\documentclass[aip,amsmath,amssymb,reprint]{revtex4-1}

\usepackage{graphicx}
\usepackage{dcolumn}
\usepackage{bm}
\usepackage{ulem}
\usepackage[utf8]{inputenc}
\usepackage[T1]{fontenc}
\usepackage{mathptmx}
\usepackage{etoolbox}
\usepackage{xcolor}
\makeatletter
\def\@email#1#2{%
 \endgroup
 \patchcmd{\titleblock@produce}
  {\frontmatter@RRAPformat}
  {\frontmatter@RRAPformat{\produce@RRAP{*#1\href{mailto:#2}{#2}}}\frontmatter@RRAPformat}
  {}{}
}%
\makeatother
\begin{document}

\preprint{}

\title[Decomposing cryptocurrency high frequency price  dynamics into recurring and noisy components]{Decomposing cryptocurrency high frequency price dynamics into recurring and noisy components}

\author{Marcin Wątorek}
\affiliation{Faculty of Computer Science and Telecommunications, Cracow University of Technology, ul.~Warszawska 24, 31-155 Krak\'ow, Poland}
\email{marcin.watorek@pk.edu.pl}

\author{Maria Skupień}
\affiliation{Department of Mathematics, Pedagogical University of Cracow, ul.~Podchorążych 2, 30-084 Krak\'ow, Poland}

\author{Jarosław Kwapień}
\affiliation{
Complex Systems Theory Department, Institute of Nuclear Physics, Polish Academy of Sciences, Radzikowskiego 152, 31-342 Kraków, Poland}

\author{Stanisław Drożdż}
\affiliation{Faculty of Computer Science and Telecommunications, Cracow University of Technology, ul.~Warszawska 24, 31-155 Krak\'ow, Poland}
\affiliation{
Complex Systems Theory Department, Institute of Nuclear Physics, Polish Academy of Sciences, Radzikowskiego 152, 31-342 Kraków, Poland}

\date{\today}

\begin{abstract}

This paper investigates the temporal patterns of activity in the cryptocurrency market with a focus on Bitcoin, Ethereum, Dogecoin, and WINkLink from January 2020 to December 2022. Market activity measures - logarithmic returns, volume, and transaction number, sampled every 10 seconds, were divided into intraday and intraweek periods and then further decomposed into recurring and noise components via correlation matrix formalism. The key findings include the distinctive market behavior from traditional stock markets due to the nonexistence of trade opening and closing. This was manifest in three enhanced-activity phases aligning with Asian, European, and U.S. trading sessions. An intriguing pattern of activity surge in 15-minute intervals, particularly at full hours, was also noticed, implying the potential role of algorithmic trading. Most notably, recurring bursts of activity in bitcoin and ether were identified to coincide with the release times of significant U.S. macroeconomic reports such as Nonfarm payrolls, Consumer Price Index data, and Federal Reserve statements. The most correlated daily patterns of activity occurred in 2022, possibly reflecting the documented correlations with U.S. stock indices in the same period. Factors that are external to the inner market dynamics are found to be responsible for the repeatable components of the market dynamics, while the internal factors appear to be substantially random, which manifests itself in a good agreement between the empirical eigenvalue distributions in their bulk and the random matrix theory predictions expressed by the Marchenko-Pastur distribution. The findings reported support the growing integration of cryptocurrencies into the global financial markets.

\end{abstract}

\maketitle

\begin{quotation}
Financial markets belong to the most complex dynamical structures associated with human activity. Their dynamics are affected by both the exogenous factors that operate on time scales largely independent of the internal market dynamics and the endogenous factors being inherent to a given market. Among the modern financial markets, the cryptocurrency market deserves special attention as it constitutes a symbol of our times and, like no other market, it operates continuously without interruptions, which allows it to react instantly to any news from the world. Based on the formalism paralleling the principal component analysis but applied to the correlation matrices connecting the consecutive trading intervals of equal length, the dynamics of price changes of several most traded cryptocurrencies is decomposed into sectors of exogenous and endogenous character. The former reflects selectively market susceptibility to the external news while the latter is largely consistent with the predictions of random matrix theory. This fact can be interpreted as an indication of the efficiency that the cryptocurrency trading dynamics has already attained.
\end{quotation}

\section{Introduction}

It is well known that financial markets have recurring patterns in their activity~\cite{Gerety1991,Andersen1998,Bollerslev2000}. There are bursts in daily volatility related to various cyclical macroeconomic reports such as NFP (Nonfarm Payrolls), CPI (Consumer Price Index), ISM (Institute for Supply Management), PMI (Purchasing Managers Index), etc.~\cite{Andersen2003,Evans2010,Evans2011,Kari2011}, central banks' monetary policy announcements~\cite{HUSSAIN2011,Kenourgios2015,gebarowski2019}, and even politician tweets~\cite{Gjerstad2021}. There is also a well-known fact that the stock market volatility shows a U-shape activity pattern related to the opening and the closing of a session~\cite{Andersen2000,Gubiec2015,Andersen2019}. An analogous intraweek effect has also been evidenced in previous research~\cite{berument2001day}. Conversely, in the foreign exchange market (Forex), which operates 24 hours a day and five days a week, periods of amplified activity can be attributed to the geographically diverse market participants~\cite{Dacorogna1993}. For example, a notable surge in activity is observed during the transitional period between the Asia/European and European/U.S. trading sessions~\cite{Zhang2018}. On the other hand, certain markets exhibit their unique patterns like, for instance, the oil market experiencing activity increase at 16:30 UTC, which coincides with the weekly release of the U.S. Energy Information Administration Petroleum Status Report on the U.S. crude oil inventories~\cite{Bu2014}. Despite these market-specific patterns, it is essential to recognize the interconnected nature of today's financial markets due to the instantaneous dissemination of information~\cite{alves2020}. For example, alterations in interest rates, primarily affecting Forex, reverberate promptly across various markets, particularly the stock market, demonstrating the profound global financial interdependence.

Since the introduction of Bitcoin in 2009, the cryptocurrency market has undergone exponential expansion, evolving from transactions primarily conducted on the internet fora to a significant market valued at over 1 trillion USD as of June 2023~\cite{coinmarket}, where transactions occur 24/7 on more than 600 platforms~\cite{coinmarket}. Despite its persistent volatility, recurrent speculation bubbles and subsequent market crashes~\cite{Gerlach2018,Bellon2022,Palomino2022,Wang2022}, security breaches~\cite{charoenwong2021decade}, and spectacular collapses~\cite{fu2022ftx,Briola2023,vidaltomas2023ftxs}, the trading characteristics of this market have been found to align considerably with those observed in the mature financial markets~\cite{bariviera2018analysis,DrozdzBTC2018,sigaki2019,watorek2021,KwapienJ-2022a,garcin2023complexity,pessa2023}. Intriguingly, the correlation between the most frequently traded cryptocurrencies, specifically bitcoin and ether, and the stock indices and other financial assets has increased significantly, which suggests an emerging interconnectedness within the global financial ecosystem~\cite{Manavi2020,Balcilar2022,Wnag2022Entr,WatorekM-2023a,Entropy2023}.

While several trading characteristics of the cryptocurrency market echo those of the regular financial markets, key differentiating aspects exist. First, the cryptocurrency market operates continuously without defined opening and closing times. Furthermore, independent exchange rates across different platforms provide opportunities for arbitrage~\cite{MAKAROV2020}. Additionally, cryptocurrency prices are unusually susceptible to all sorts of rumors, like tweets~\cite{Ante2023}, and manipulations in the form of the pump-and-dump schemes~\cite{li2021cryptocurrency,Dhawan2022} and wash trading~\cite{cong2022crypto}. It is worth noting that despite a large universe of more than 10,000 cryptocurrencies, the market is largely dominated by bitcoin and ether. As of June 2023, these two assets represent approximately 70\% of the market's total capitalization, of which bitcoin accounts for around 50\% and ether accounts for around 20\%~\cite{coinmarket}.

Even with a relatively nascent status of the cryptocurrency market, some research investigating its intraday and intraweek trading patterns has already been conducted. A time-of-day, day-of-week, and month-of-year periodicity in the bitcoin volatility and liquidity have been reported~\cite{BAUR2019} as well as a marked decrease in trading activity during weekends~\cite{BAUR2019,KwapienJ-2022a}. Moreover, atypical trade intensity and bitcoin volatility on Thursdays and Fridays have been identified~\cite{Catania2019}. Linkages between the amplified bitcoin trading activity and the trading hours of the major global stock markets have also been observed~\cite{Dyhrberg2018,Eross2019,Wang2020FRL}. Some influence of macroeconomic news on the intraday seasonal volatility for bitcoin and ether in the Gemini exchange before 2020 has also been found~\cite{Omrane2023}. Periodicity differences between the bitcoin and ether volatility and their liquidity across centralized (e.g., Binance and Coinbase) and decentralized (e.g., Uniswap) exchanges have been reported~\cite{Hansen2022}. In the centralized exchanges, the patterns observed in high-frequency data were attributed to algorithmic trading~\cite{Hansen2022}. However, earlier studies showed no effect of algorithmic trading on the intraday patterns of cryptocurrencies~\cite{Petukhina2021}. Some studies have explored the microstructure characteristics of the bitcoin spot and futures markets~\cite{Aleti2021} and identified a higher degree of the microstructure noise in the cryptocurrency market in comparison to stock markets~\cite{DIMPFL2021}.

Given the above observations, one may ask whether there exist recurrent patterns in the cryptocurrency market volatility and trading activity that parallel those observed in the traditional financial markets or they are fundamentally dissimilar. The present contribution addresses this question by employing the correlation matrix formalism and signal decomposition into principal components to isolate recurring activity patterns from noise, based on recent high-frequency price changes, volume, and the number of transactions. This approach has previously been proven successful in revealing patterns in the German stock index DAX~\cite{Drozdz2001PhysA} and the brain sensory response~\cite{Kwapien2000}, as well as in capturing collectivity in the stock~\cite{Drozdz2000,James2021chaos,James2022PhysAstock} and cryptocurrency~\cite{Chaudhari2020,Chaos2020,James2022,Nguyen2022,gavin2023community,James2023} markets together with its temporal changes in the cryptocurrency market~\cite{entropy2021b,JAMES2022PhysD}.

\section{Data and methods}

\subsection{Data description and properties}

The data set considered in this study encompasses four cryptocurrencies: Bitcoin (BTC), Ether (ETH), Dogecoin (DOGE), and WINkLink (WIN) and three observables: logarithmic price changes, transaction volume, and the number of transactions in time unit. Time series spanning 3 years (Jan 1, 2020 to Dec 31, 2022) have been downloaded from the Binance exchange~\cite{Binance}, which offers price quotations recorded with 10-second frequency. All prices are denominated in tether (USDT), the most frequently traded stablecoin on Binance.

BTC and ETH are the two highest capitalized and most liquid cryptocurrencies with the average inter-transaction time in the considered period being $\delta t_{\rm BTC}=0.04s$ and $\delta t_{\rm ETH}=0.1s$~\cite{Entropy2023}). The price changes of these cryptocurrencies are strongly correlated with each other~\cite{watorekfutnet2022} and, as it has recently been documented~\cite{Entropy2023}, with the U.S. stock indices. DOGE is the most influential one among the meme coins; its spectacular price movements attracted media attention due to the impact of Elon Musk's tweets~\cite{Nani2022}. Despite its smaller capitalization compared to BTC and ETH, DOGE maintains substantial liquidity on the Binance exchange ($\delta t_{\rm{DOGE}}=0.2s$~\cite{Entropy2023}). WIN, a smart contract oracle and competitor to Chainlink, boasts significantly smaller capitalization, ranking beyond the top 100 cryptocurrencies~\cite{coinmarket}. However, its inclusion here is justified by its distinct characteristics: absence of a correlation with the U.S. stock indices~\cite{Entropy2023}. At the same time, its average inter-transaction time ($\delta t_{\rm{WIN}}=1.1s$~\cite{Entropy2023}) is sufficient to carry out the analysis from a statistical perspective.

The logarithmic returns, derived from the price changes $p(t)$, were calculated as $R_{\Delta t}(t)=\textrm{log}p(t)-\textrm{log}p(t-\Delta t)$, where $\Delta t=10$s. Subsequently, in conjunction with transaction volumes $V_{\Delta t}$ and the number of transactions $N_{\Delta t}$, the return time series were segmented into day-long and week-long intervals. The former, ranging from 00:00:00 to 23:59:50, comprised $T_{\rm{day}}=8,640$ data points across $K_{\rm{day}}=1,096$ days. The latter, ranging from Sunday 00:00:00 to Saturday 23:59:50, comprised $T_{\rm{week}}=60,480$ data points across $K_{\rm{week}}=156$ weeks. The in-segment time series $R_{d}(i)$, $V_{d}(i)$, $N_{d}(i)$ represent daily trading ($d=1,...,K_{\rm{day}}$ and $i=1,...,T_{\rm{day}})$) and $R_{w}(j)$, $V_{w}(j)$, $N_{w}(j)$ represent weekly trading ($w=1,...,K_{\rm{week}}$ and $j=1,...,T_{\rm{week}}$).

\begin{figure*}
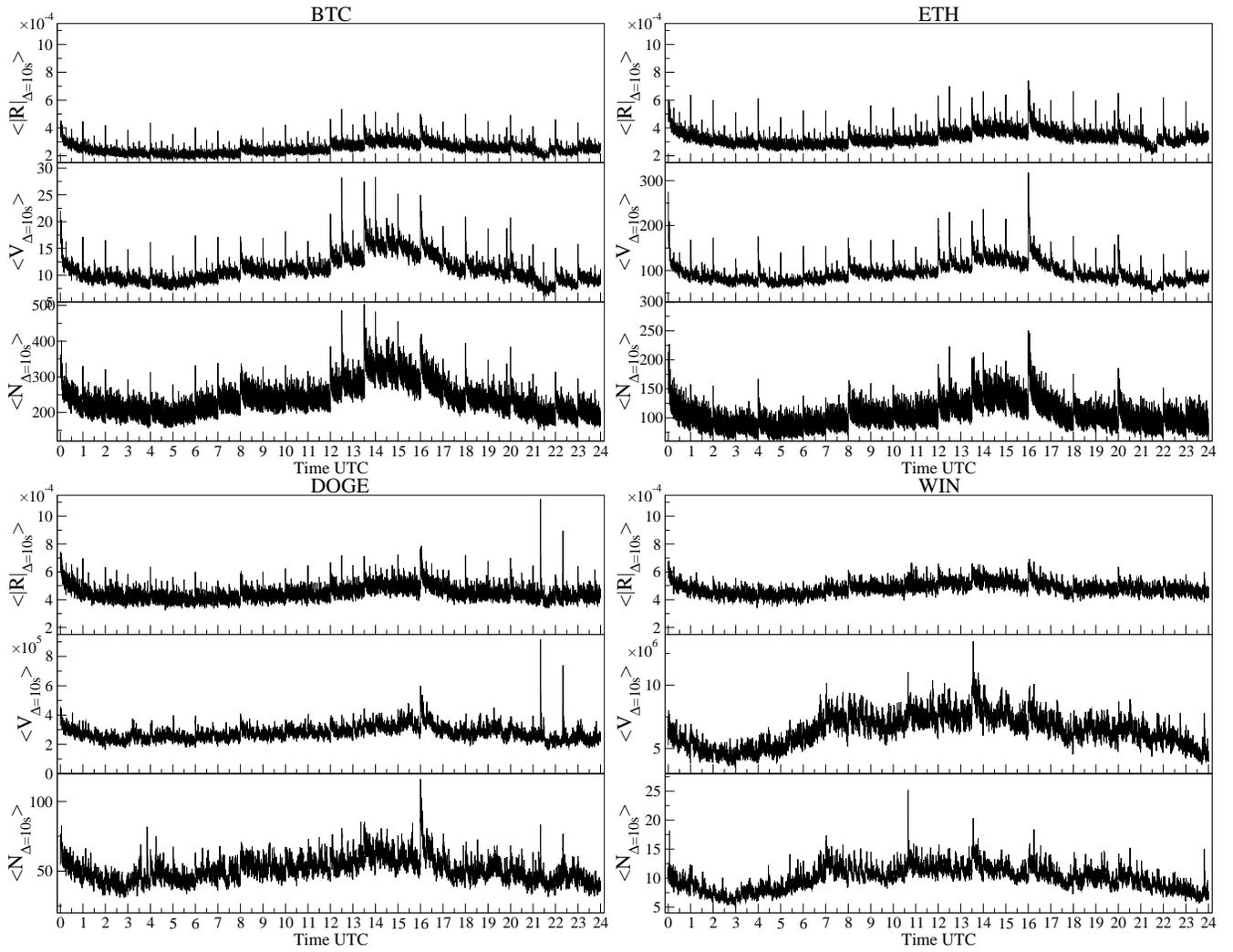

\includegraphics[width=0.49\textwidth]{figs/24havarageBTC.eps}
\includegraphics[width=0.49\textwidth]{figs/24havarageETHu.eps}
\includegraphics[width=0.49\textwidth]{figs/24havarageDOGEu.eps}
\includegraphics[width=0.49\textwidth]{figs/24havarageWINu.eps}
\caption{Intraday time series of average volatility $|R_{\Delta t}|$, volume $V_{\Delta t}$, and the transaction number $N_{\Delta t}$ sampled with $\Delta t=10$s for BTC, ETH, DOGE, and WIN. Note that the y-scale is different in the case of $N$ and $V$.}
\label{fig::Average24}
\end{figure*}

\begin{figure}
\includegraphics[width=0.49\textwidth]{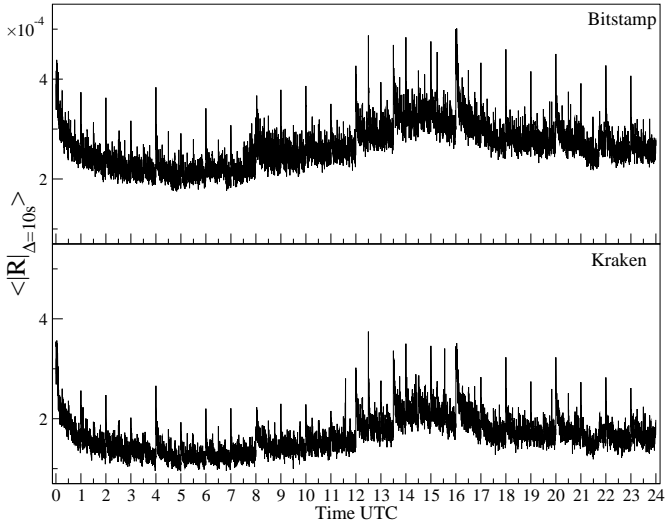}
\caption{Intraday time series of average volatility $|R_{\Delta t}|$ sampled with $\Delta t=10$s for BTC on Bitstamp (top) and Kraken (bottom).}
\label{fig::avargeKR_BI}
\end{figure}

Initial patterns of the trading activity can be inferred from the average values of $|R_{\Delta t}|$, $V_{\Delta t}$, and $N_{\Delta t}$ calculated across the trading days, as illustrated in Fig.~\ref{fig::Average24}. Predominantly, liquidity disparities assessed through the number of transactions $N_{\Delta t}$ and volatility $|R_{\Delta t}|$ become apparent. As expected, BTC that is the highest capitalized and most recognizable cryptocurrency registers the most transactions per 10-second interval. Comparable transaction numbers are observed for ETH, particularly when accounting for its higher volume (in ETH units) and lower price. Conversely, DOGE exhibits on average half as many transactions per 10-second interval as ether. In the case of WIN a value of $N_{\Delta t}$ oscillates around 10.
\begin{figure}
\includegraphics[width=0.49\textwidth]{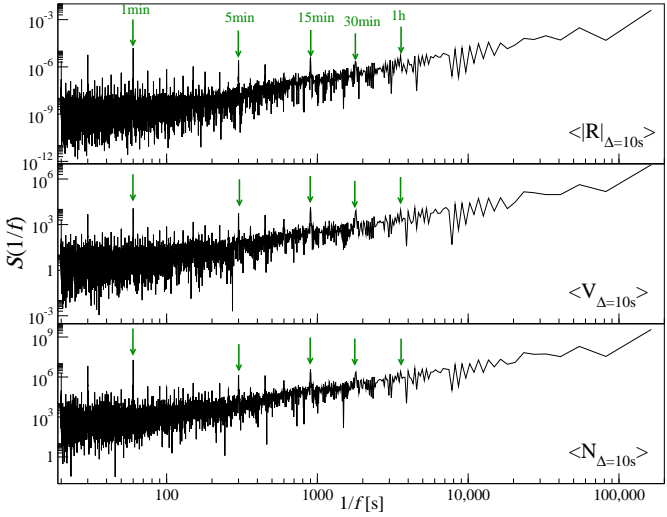}
\caption{Power spectral density $S(1/f)$ of intraday time series presented in Fig.~\ref{fig::Average24} for BTC. Note that in order to better reflect the time relations $1/f$ is used as an argument. Also the vertical axes scale is different in each of the panels.}
\label{fig::PSD_BTC24}
\end{figure}
A strong pattern is visible in Fig.~\ref{fig::Average24} in a form of the periodic spikes every full hour with minor peaks occurring every half an hour and quarter an hour for BTC and ETH. This pattern consists of increased volatility, volume, and the number of transactions. The related variations are also visible in the frequency domain, as illustrated in Fig.\ref{fig::PSD_BTC24}. A similar pattern is also present for DOGE, albeit the hourly increase of volume and the number of transactions is less pronounced here. In contrast, such a pattern is absent in the least liquid of the considered cryptocurrencies, WIN. A potential explanation for the observed periodic patterns can be algorithmic trading, the effects of which have already been documented in stock markets as the periodic activity patterns corresponding to minutes~\cite{Muravyev2022} and hours~\cite{Broussard2014}. Interestingly, the comparable full-hour patterns have recently been reported for cryptocurrencies~\cite{Hansen2022}. The same phenomenon of the amplified volatility $|R_{\Delta t}|$ at every full quarter, every half an hour, and every full hour can be observed during the same period (2020-2022) on the Kraken and Bitstamp exchanges as it is demonstrated in Fig.~\ref{fig::avargeKR_BI}.

Different fluctuation patterns are associated with the trading activity of individual users on Binance, which is currently the largest cryptocurrency exchange in terms of the traded volume value~\cite{coinmarket}. The Binance users are globally distributed, with the largest groups of users, each around 5\%, residing in Turkey, Russia, Argentina, Vietnam, and Ukraine, while the remaining 75\% are located in other countries~\cite{similarweb}. Consequently, the peak volatility, volume, and transaction numbers are observed between 12:00 and 16:00 UTC, a timeframe coinciding with the overlap of the European and American trading sessions. Following this period, there is a significant decrease in market activity, reaching a day low after 21:00 UTC with the closure of the U.S. trading session. Another surge in the market activity aligns with the onset of the Asian trading session at 00:00 UTC, followed by a subsequent decline until the opening of the European session at 06:00 UTC. This pattern is particularly evident for the most liquid cryptocurrencies, BTC and ETH. However, this pattern is slightly perturbed for DOGE and apart from the peaks at 00:00 and 16:00 UTC, two additional peaks are noted at 21:20 and 22:20 UTC, which correspond to the highest volatility and volume periods. The least liquid one of the considered cryptocurrencies, WIN, exhibits near-constant volatility over a 24-hour period. The diurnal volume and transaction number patterns align with those observed for the most liquid cryptocurrencies.

\begin{figure*}
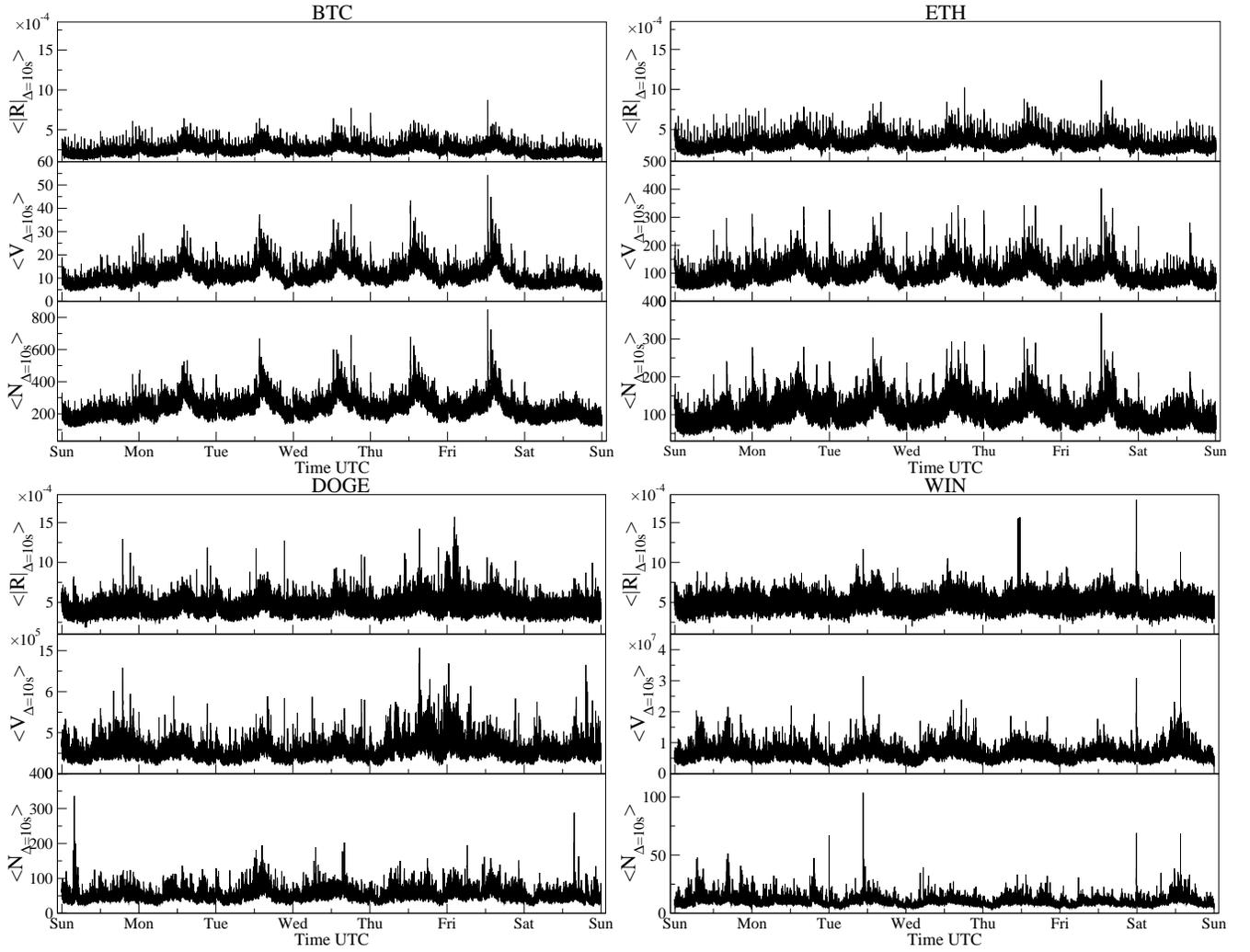

\includegraphics[width=0.49\textwidth]{figs/weekBTCavarage.eps}
\includegraphics[width=0.49\textwidth]{figs/weekETHavarageu.eps}
\includegraphics[width=0.49\textwidth]{figs/weekDOGEavarageu.eps}
\includegraphics[width=0.49\textwidth]{figs/weekWINavarageu.eps}
\caption{Intraweek time series of average volatility $|R_{\Delta t}|$, volume $V_{\Delta t}$, and the transaction number $N_{\Delta t}$ sampled with $\Delta t=10$s for BTC, ETH, DOGE, and WIN.}
\label{fig::Averageweek}
\end{figure*}

Distinct weekly patterns are also discernible in the cryptocurrency market where trading is daily, as it is demonstrated in Fig.~\ref{fig::Averageweek}. For BTC and ETH, conspicuous surges of activity are seen during weekdays at midday, coinciding with the overlap of the European and American trading sessions. The largest values of $|R_{\Delta t}|$, volume $V_{\Delta T}$, and the transaction number $N_{\Delta t}$ manifest themselves on Fridays. The market dynamics on Thursdays, Wednesdays, and Tuesdays exhibit similar trends. Unlike this, a slight attenuation of trading activity is observed on Mondays. During weekends, when the traditional financial markets are closed, the cryptocurrency market exhibits significantly calmer trading. No such regular weekly pattern can be observed for the less liquid DOGE and WIN.

\subsection{Correlation matrix and random matrix limit}

After the exploration of the cryptocurrency market dynamics via the comparison of the average intraday and intraweek values, the correlation matrix formalism will be used now to calculate the intraday correlations in market activity. The most recurrent dynamical structures will be captured through the eigenvalues decomposition. Correlation matrix is defined as ${\bf C}^X_{\rm{P}} = {1 \over T_{\rm{P}}} {\bold M}^X_{\rm{P}} {\tilde {\bold M}}_{\rm{P}}^X$, where ${\bold M}^X_{\rm{P}}$ is a data matrix of size $K_{\rm{P}} \times T_{\rm{P}}$ and $\tilde{\cdot}$ denotes matrix transpose. In this general notation, $X$ stands for $R_{\Delta t}$, $V_{\Delta t}$, or $N_{\Delta t}$, whereas P is day or week. After diagonalization of the correlation matrix ${\bf C}^X_{\rm{P}}$
\begin{equation}
{\bf C}^X_{\rm{P}} {\bf v}_k = \lambda_k {\bf v}_k,
\label{eigenequation}
\end{equation}
its eigenvalues $\lambda_k$ are derived alongside the corresponding eigenvectors ${\bf v}_k=\{v_{km}\}$, where $k=1,...,K_{\rm{P}}$. Here, $m \in [1,T_{\rm{day}}]$ applies to intraday dynamics and $m \in [1,T_{\rm{week}}]$ applies to intraweek dynamics.

The derived empirical eigenvalue distribution can be tested against the Marchenko-Pastur distribution corresponding to the Wishart ensemble of random matrices ${\bf W}$~\cite{randommatrix} that represent the universal properties of uncorrelated i.i.d. random variables with the Gaussian distribution $N(0,\sigma)$\cite{wishart1928}. The probability density function defining an eigenvalue distribution of $\bf W$ has the analytical form
\begin{equation}
\phi_\textrm{W}(\lambda)={1 \over N} \sum_{k=1}^N \delta(\lambda - \lambda_k) = {Q \over 2 \pi \sigma_{\textrm{W}}^2} {\sqrt{(\lambda_{+}-\lambda)(\lambda-\lambda_{-})} \over \lambda},
\label{rhoW}
\end{equation}
\begin{equation}
\lambda_{\pm} = \sigma_{\textrm{W}}^2 (1 + 1/Q \pm 2 \sqrt{1 \over Q}),
\label{lambdaW}
\end{equation}
where $\lambda\in[\lambda_-,\lambda_+]$ and $Q=T_{\rm P}/K_{\rm P}$. This relationship is strictly valid in the infinite limit $T_{\rm P}, K_{\rm P} \to \infty$\cite{Marchenko1967} but a comparison of the empirical eigenvalue distribution with the Marchenko-Pastur distribution helps distinguish whether any correlated structures exist in the data. In the present case one may look for repeatable intraday or intraweek structures.


\section{Intraday patterns}

\subsection{Correlation characteristics}

\begin{figure*}
\includegraphics[width=0.99\textwidth]{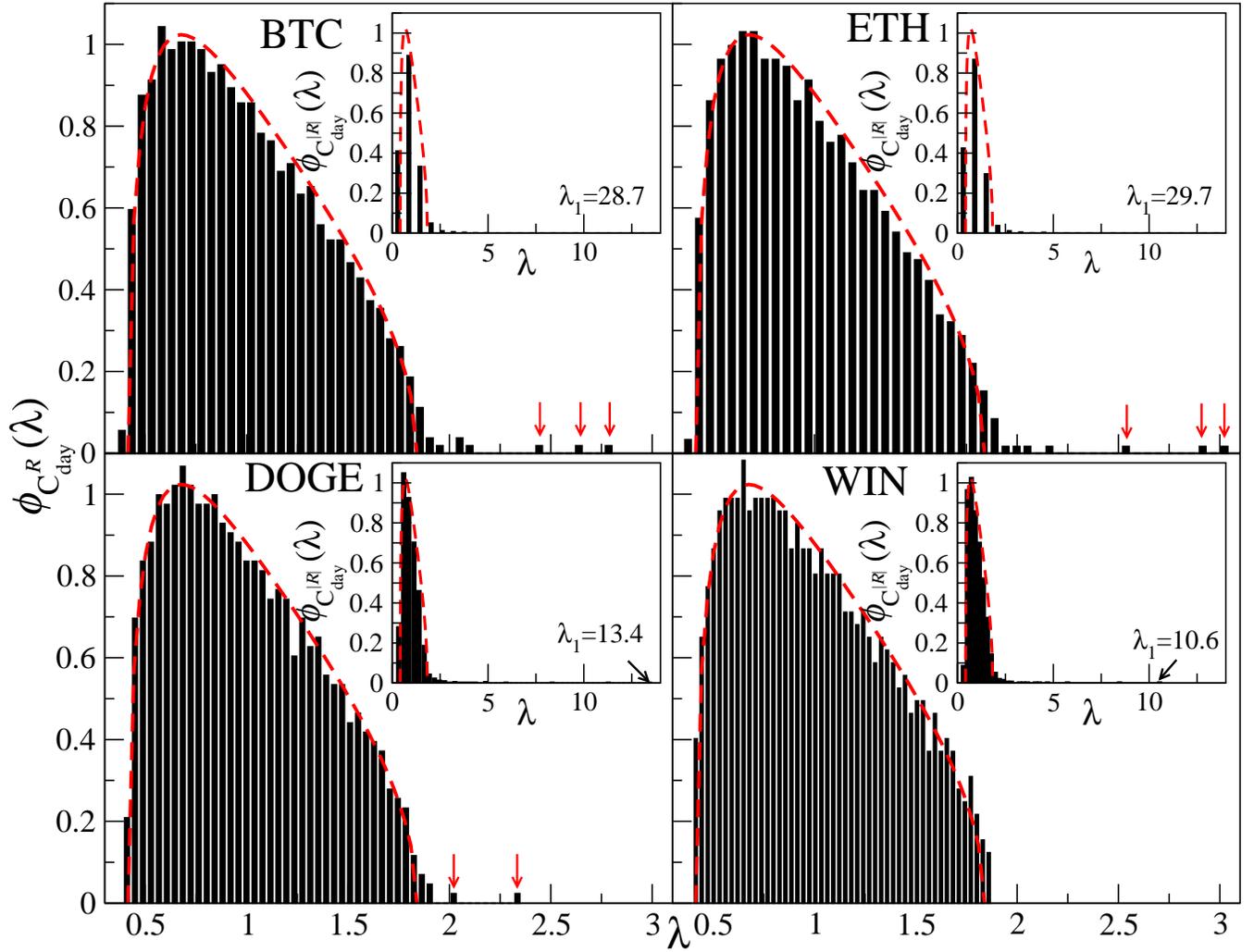}
\caption{Eigenvalue distribution $\phi^R_{\textrm{day}}$ obtained from the correlation matrix of intraday log-returns ${\bf C}^R_{\rm{day}}$ for BTC, ETH, DOGE, and WIN. Insets show the eigenvalue distributions $\phi^{|R|}_{\textrm{day}}$ obtained from the correlation matrix of absolute intraday log-returns ${\bf C}^{|R|}_{\rm{day}}$. Eigenvalues outside the Marchenko-Pastur region (dashed red lines) have been marked.}
\label{fig::MPdistrall24hR}
\end{figure*}

\begin{figure}
\includegraphics[width=0.49\textwidth]{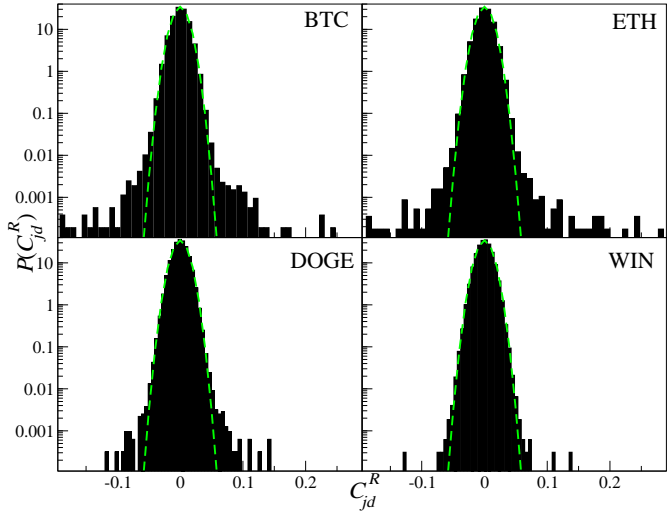}
\caption{Probability density function of intraday off-diagonal elements of the correlation matrix ${\bf C}^R_{\rm{day}}$ for BTC, ETH, DOGE and WIN. Dashed line shows a fitted normal distribution.}
\label{fig::Rozklelmac24R}
\end{figure}

In this section, the properties of correlation matrices ${\bf C}^{R}_{\rm{day}}$, ${\bf C}^{V}_{\rm{day}}$ and ${\bf C}^{N}_{\rm{day}}$ are explored. The eigenvalue distributions are juxtaposed with the theoretical Marchenko-Pastur distribution for random matrices to assess the presence of intraday structures that are genuinely recurring. As depicted in Fig.\ref{fig::MPdistrall24hR}, the majority of the empirical eigenvalues for $R_{\Delta t}$ reside within the Marchenko-Pastur (M-P region). Specifically, all eigenvalues fall within this region in the case of WIN. This indicates that there is no correlation among the intraday returns for this cryptocurrency, which implies no recurring intraday structure is detectable. The random nature of the correlations is further validated by the nearly Gaussian form of the off-diagonal element distribution for WIN, as shown in Fig.\ref{fig::Rozklelmac24R} (although even in this case the Gaussianity of the distribution is still rejected by Kolmogorov-Smirnov and Jarque-Bera tests). This suggests that the cross-correlations are essentially random. A situation is different for the most liquid cryptocurrencies: BTC and ETH. Here, a handful of the eigenvalues are found outside the random region, signifying the existence of distinct, time-specific, recurrent structures in the intraday trading. The off-diagonal element distributions for BTC and ETH, illustrated in Fig.~\ref{fig::Rozklelmac24R}, deviate distinctly from the Gaussian distribution. Contrary to the situation with WIN, here the p.d.f. tails are heavier. DOGE presents a scenario intermediate to the previous ones with only two eigenvalues situated clearly outside the M-P region. This hints at the existence of some repetitive patterns, albeit less pronounced than in the case of BTC and ETH.

It is noteworthy that for all the considered cryptocurrencies, the eigenvalue bulk for $R_{\Delta t}$ aligns well with the bounds defined by the Marchenko-Pastur distribution for random matrices. This observation underscores the lack of shared information among the respective log-returns in different trading days. Conversely, multiple eigenvalues are situated outside the M-P region in the case of volatility $|R|$ (see the insets in Fig.\ref{fig::MPdistrall24hR}), which indicates recurrence in $|R_{\Delta t}|$. This is a manifestation of the well-documented stylized facts demonstrating that there is no autocorrelation in log-returns and a long (which decays as a power law) memory in volatility~\cite{Gopikrishnan1999,Ausloos2000,ContR-2001a,jiang2019multifractal,Klamut2020,Klamut2021}.

\begin{figure}
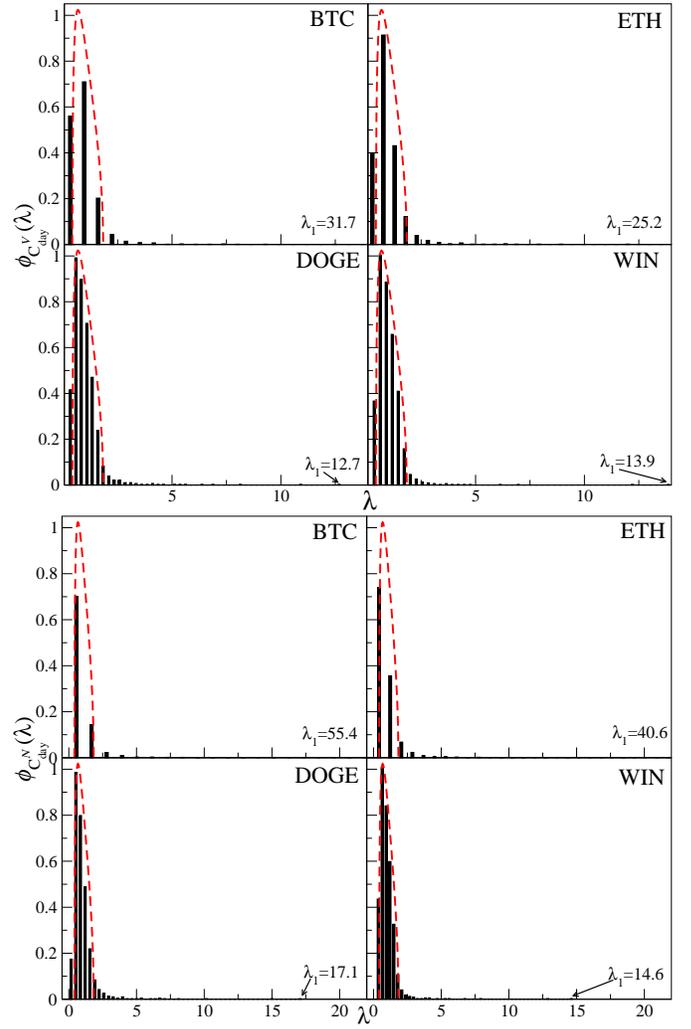

\includegraphics[width=0.49\textwidth]{figs/MPdistr24hVu.eps}
\includegraphics[width=0.49\textwidth]{figs/MPdistr24hNu.eps}
\caption{Eigenvalue distribution $\phi^V_{\textrm{day}}$ and $\phi^N_{\textrm{day}}$ obtained from the correlation matrix of intraday volume ${\bf C}^V_{\rm{day}}$ (upper) and the intraday number of transactions ${\bf C}^N_{\rm{day}}$ (lower) for BTC, ETH, DOGE and WIN. In each case, the Marchenko-Pastur distribution is shown (dashed red lines) and the value of $\lambda_{1}$ is listed.}
\label{fig::MPdistrall24hVN}
\end{figure}

\begin{figure}
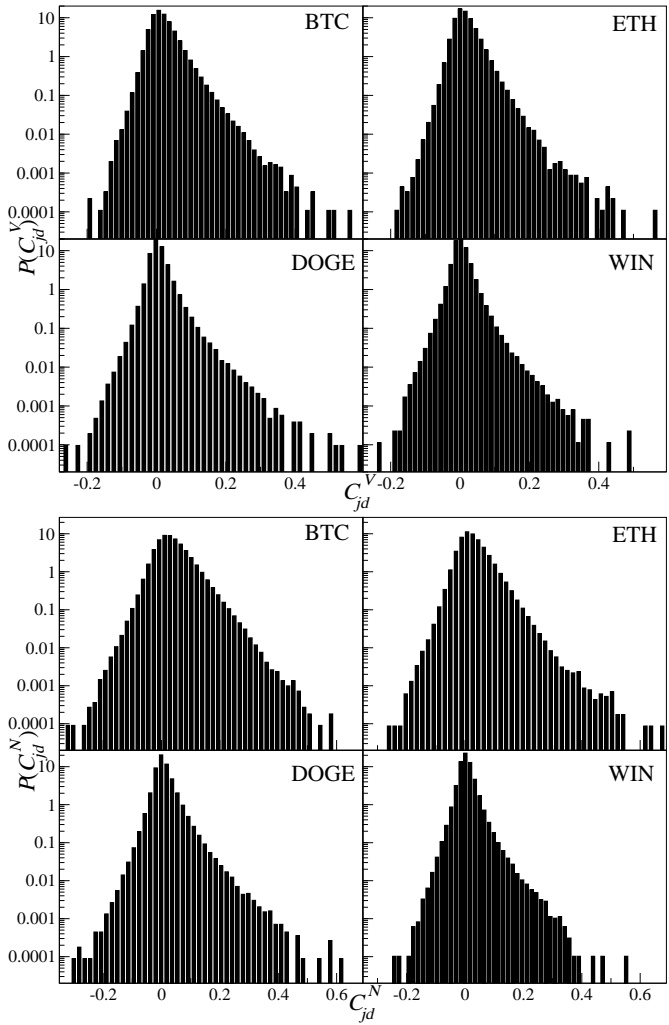

\includegraphics[width=0.49\textwidth]{figs/Rozklelmac24hV.eps}
\includegraphics[width=0.49\textwidth]{figs/Rozklelmac24hN.eps}
\caption{Probability density function of the off-diagonal elements for the intraday correlation matrix ${\bf C}^V_{\rm{day}}$ (upper) and ${\bf C}^N_{\rm{day}}$ (lower) for BTC, ETH, DOGE, and WIN.}
\label{fig::Rozklelmac24VN}
\end{figure}

The correlations between $V_{\Delta t}$ and $N_{\Delta t}$ have different properties. They display a higher level of collectivity compared to $R_{\Delta t}$. A significant number of eigenvalues deviate notably from the random region (as shown in Fig.~\ref{fig::MPdistrall24hVN}) and the off-diagonal element distributions (as shown in Fig.\ref{fig::Rozklelmac24VN}) also deviate substantially from the Gaussian distribution. Such effects are known to induce the eigenvalues standing outside the random matrix theory regime~\cite{Drozdz2002RMT}. The global collective behavior epitomized by the largest eigenvalue $\lambda_1$ is more pronounced in the case of BTC and ETH. These observations suggest that the intraday structures for volatility $V_{\Delta t}$ and the transaction number $N_{\Delta t}$ recur more frequently than those for log-returns $R_{\Delta t}$. Furthermore, these findings corroborate the existence of repeatable patterns of correlated Binance user activity already seen in the average volume and transaction number in Fig.~\ref{fig::Average24}.

Another noteworthy observation is related to the congruence between the eigenvalue distributions and $\lambda_1$ for volatility $|R_{\Delta t}|$ and volume $V_{\Delta t}$. It suggests that the cross-correlations between the intraday absolute price fluctuations and the intraday volume are similar, implying that both quantities share the intraday patterns. This correspondence between volume and volatility can be interpreted as a manifestation of a linear form of the price impact function~\cite{BouchaudJP-2010a}. This particular relationship form has been subject of recent research focusing on the properties of the cryptocurrency market~\cite{Entropy2023}.

\subsection{Superposed time series}

\begin{figure}
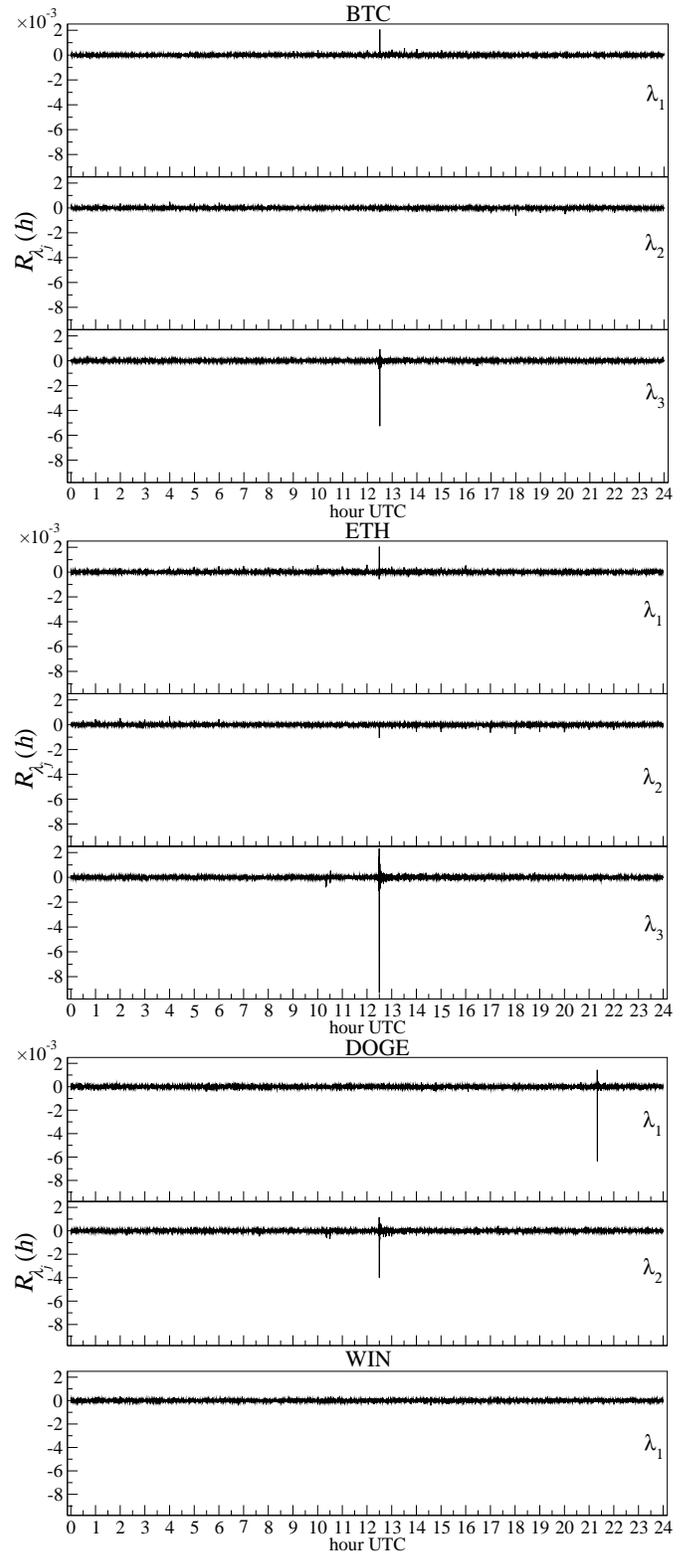

\includegraphics[width=0.49\textwidth]{figs/24R_BTC3.eps}
\includegraphics[width=0.49\textwidth]{figs/24R_ETH3.eps}
\includegraphics[width=0.49\textwidth]{figs/24R_DOGE_WIN.eps}
\caption{The superposed time series of log-returns $R_{\lambda_k}(i)$ calculated according to Eq.~(\ref{eigensignalday}) for BTC, ETH, DOGE, and WIN.}
\label{fig::eigR24}
\end{figure}

From an investor perspective, the most compelling facet of the market dynamics pertains to the timing of the recurrent patterns of the amplified market activity. To identify such intervals, the superposed time series are constructed in a form of eigensignals~\cite{Drozdz2000}:
\begin{equation}
R_{\lambda_k}(i)= \sum_{d=1}^{K_{\rm{day}}} {\rm sign}(v_{kd})|v_{kd}|^2 R_{\Delta t}(i).
\label{eigensignalday}
\end{equation}
As illustrated in Fig~\ref{fig::MPdistrall24hR}, there are three eigenvalues notably outside the random region for the log-returns of BTC and ETH. Two such eigenvalues are present for DOGE, whereas none exists for WIN. These eigenvalues represent non-random intraday correlations and, hence, the corresponding eigenvectors might contain valuable insights into the recurrent structures of market dynamics. Consequently, the superposed time series for $k=1,2,3$ in the case of BTC and ETH and $k=1,2$ for DOGE are presented in Fig.\ref{fig::eigR24}. Although all eigenvalues for WIN are located in the M-P region, the superposed time series for $k=1$ are also presented for comparison. The superposed log-returns $R_{\lambda_1}$ for the least liquid cryptocurrency WIN merely represents noise (as anticipated). On the other hand, the most collective signal associated with $\lambda_1$ portrays the most profound synchronicity of BTC and ETH, which occurs around 12:30 UTC. At this time, the numerous U.S. economic data are published. This timing also coincides with the peak synchronous activity reported for the DAX stock index\cite{Drozdz2001PhysA}. The second eigenvalue $\lambda_2$ reflects the periodic market activity corresponding to full hours, potentially instigated by algorithmic trading. However, this pattern is significantly weaker than the one linked to $\lambda_1$. The third eigensignal $R_{\lambda_3}$ also pinpoints the timeframe around 12:30 UTC. In contrast to $\lambda_1$, it points out to the occurrence of negative log-returns during that period.

Different patterns are observable in a signal corresponding to the most collective eigenvector for DOGE. In this case, the most frequent patterns in log-returns occur at 21:20 UTC. Notwithstanding, the same timeframe as in the case of $\lambda_1$ for BTC and ETH, i.e., around 12:30 UTC, is indicated by $R_{\lambda_2}$. One plausible explanation for the disparate recurring patterns observed in DOGE could be the influence of Elon Musk's tweets and the subsequent volatility spikes following their dissemination, as it was discussed in~\cite{SHAHZAD2022Doge}.

In the superposed log-return time series $R_{{\Delta t}_k}$, the peak synchronous activity was observed to occur at 12:30 UTC for BTC and ETH. This raises a question which events occur on the most correlated days during that time. To address it, the off-diagonal elements of the correlation matrix ${\bf C}^{R}_{\rm{day}}$ for BTC are arranged in descending order. These elements, along with the timing of the pattern and the releases of macroeconomic news, are presented in Tab.~\ref{tab::macro}. Consistent with the superposed time series evolution associated with the most collective eigenvector ${\bf v}_1$, the increased activity for nearly all strongly correlated days was found to happen at 12:30 UTC. Several U.S. macroeconomic news are released at this time on varying weekdays. The most recurrent events are the NFP - Nonfarm Payrolls report on unemployment (Friday 12:30) and the reports related to inflation: CPI - Consumer Price Index, PPI - Producer Price Index, PCE - Personal Consumption Expenditures, together with PI - Personal Income (Tue-Fri 12:30). The only exceptions are the two days featuring the publication of the statement post Federal Reserve meetings on Wednesday at 18:00. No day exhibiting a strong correlation has been found on weekends, which constitute periods devoid of any macroeconomic announcement in the U.S., and also on Monday, which is generally a quiet day for economic news. This suggests a degree of correlation between patterns in the BTC market activity and the news related to the U.S. economy. Interestingly, while the recurring patterns on the most strongly correlated days occurred at the consistent time of 12:30, these patterns emerged on different weekdays. Hence, a further investigation of the intraweek patterns of market dynamics would be appropriate.

\begin{table}[]
\caption{The largest off-diagonal elements of the correlation matrix $C^{R}_{\rm{day}}$ for BTC, together with the time and date of the U.S. macroeconomic news releases that were related to the increased market activity. NFP - Nonfarm Payrolls, CPI -Consumer Price Index, PPI - Producer Price Index, PCE - Personal Consumption Expenditures, Pi Personal Income, FED - FOMC (Federal Reserve Board and Federal Open Market Committee) statement.} 
\label{tab::macro}
\footnotesize
\begin{tabular}{|c|c|c|c|c|c|c|c|c|}
\hline
\textbf{Date} & \textbf{time} & \textbf{day} & \textbf{event} & \textbf{$C^{R}_{jd}$} & \textbf{Date} & \textbf{time} & \textbf{day} & \textbf{event} \\ \hline
07.10.2022 & 12:30 & FRI & NFP & 0.25 & 28.10.2022 & 12:30 & FRI & PCE+PI \\ \hline
21.09.2022 & 18:00 & WED & FED & 0.23 & 03.11.2021 & 18:00 & WED & FED \\ \hline
07.10.2022 & 12:30 & FRI & NFP & 0.23 & 12.10.2022 & 12:30 & WED & PPI \\ \hline
12.10.2022 & 12:30 & WED & PPI & 0.18 & 28.10.2022 & 12:30 & FRI & PCE+PI \\ \hline
07.10.2022 & 12:30 & FRI & NFP & 0.17 & 04.11.2022 & 12:30 & FRI & NFP \\ \hline
13.09.2022 & 12:30 & TUE & CPI & 0.14 & 13.07.2022 & 12:30 & WED & CPI \\ \hline
15.11.2022 & 12:30 & TUE & PPI & 0.13 & 13.12.2022 & 12:30 & TUE & CPI \\ \hline
10.11.2022 & 12:30 & THU & CPI & 0.12 & 13.12.2022 & 12:30 & TUE & CPI \\ \hline
10.06.2022 & 12:30 & FRI & NFP & 0.12 & 07.10.2022 & 12:30 & FRI & NFP \\ \hline
12.10.2022 & 12:30 & WED & PPI & 0.12 & 04.11.2022 & 12:30 & FRI & NFP \\ \hline
10.06.2022 & 12:30 & FRI & NFP & 0.12 & 13.09.2022 & 12:30 & TUE & CPI \\ \hline
13.07.2022 & 12:30 & WED & CPI & 0.12 & 30.09.2022 & 12:30 & FRI & PCE \\ \hline
10.06.2022 & 12:30 & FRI & NFP & 0.12 & 13.07.2022 & 12:30 & WED & CPI \\ \hline
10.06.2022 & 12:30 & FRI & NFP & 0.11 & 13.10.2022 & 12:30 & THU & CPI \\ \hline
13.10.2022 & 12:30 & THU & CPI & 0.11 & 13.09.2022 & 12:30 & TUE & CPI \\ \hline
12.10.2022 & 12:30 & WED & PPI & 0.11 & 13.10.2022 & 12:30 & THU & CPI \\ \hline
28.10.2022 & 12:30 & FRI & PCE+PI & 0.10 & 04.11.2022 & 12:30 & FRI & NFP \\ \hline
07.10.2022 & 12:30 & FRI & NFP & 0.10 & 05.08.2022 & 12:30 & FRI & NFP \\ \hline
\end{tabular}
\end{table}

\section{Intraweek patterns}

\begin{figure}
\includegraphics[width=0.49\textwidth]{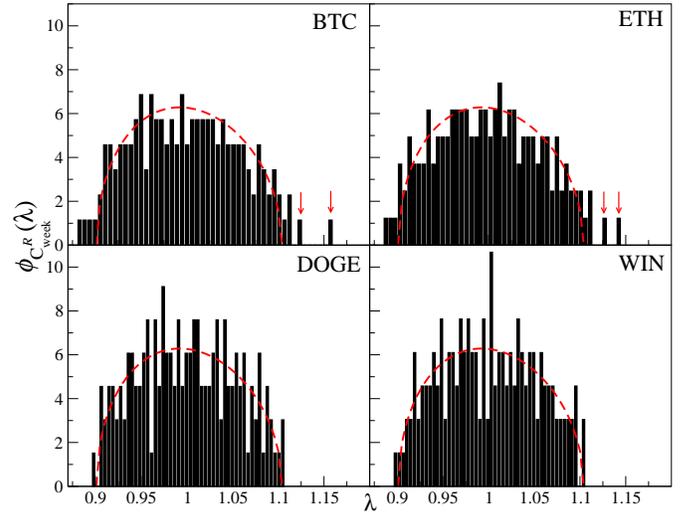}
\caption{Eigenvalue distribution $\phi^R_{\textrm{week}}$ obtained from the correlation matrix of intraweek log-returns ${\bf C}^R_{\rm{week}}$ for BTC, ETH, DOGE and WIN. In each case, the Marchenko-Pastur distribution is shown (dashed red lines) and the eigenvalues outside it have been marked.}
\label{fig::MPdistrweekR3}
\end{figure}

In the preceding section, the existence of recurring structures within the intraday dynamics of the cryptocurrency market at specific temporal intervals was identified. It can be read from Table \ref{tab::macro} that formation of these patterns is typically influenced by particular weekdays and the U.S. economic news releases. However, when analyzing it from a daily decomposition perspective, it is not feasible to distinguish the days exhibiting the strongest synchronous activity. In order to quantify this, the intraweek dynamics needs to be investigated as well by constructing the correlation matrix ${\bf C}^{R}_{\rm{week}}$ and the corresponding superposed time series:
\begin{equation}
R_{\lambda_k}(j)= \sum_{w=1}^{K_{\rm{week}}} {\rm{sign}} (v_{kw})|v_{kw}|^2 R_{\Delta t}(j).
\label{eigensignalweek}
\end{equation}
The eigenvalue distribution $\phi^R_{\textrm{week}}$ is presented in Fig.~\ref{fig::MPdistrweekR3} to expose the patterns inherent to the intraweek market activity. Upon comparing it with the eigenvalue distribution $\phi^R_{\textrm{day}}$ for the intraday correlation matrix (Fig.~\ref{fig::MPdistrall24hR}), a lower number of eigenvalues is observed outside the M-P region. This result indicates that the collective behaviour is weaker here. It has been expected, however, given that the recurrent patterns are spread over an extended span of days. In the case of DOGE and WIN, all eigenvalues are contained within the M-P region, thereby confirming the absence of any recurrent intraweek structure. In contrast, three eigenvalues are detected outside the M-P distribution for BTC and ETH, leading to the possible computation of the superposed time series $R_{\lambda_k}(j)$ for $k=1,2,3$.

\begin{figure}
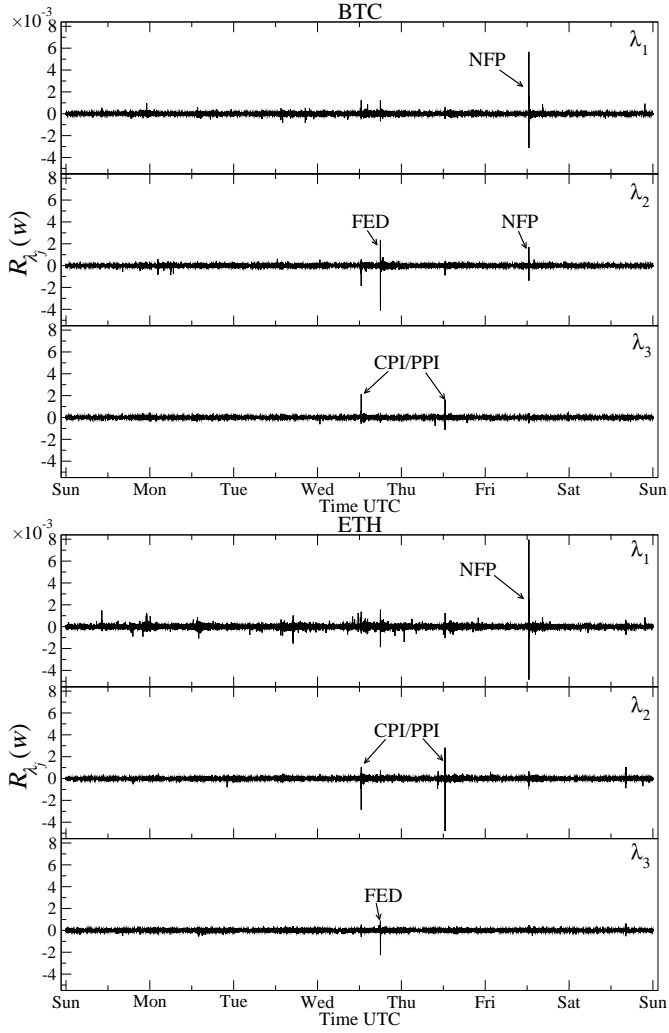

\includegraphics[width=0.49\textwidth]{figs/BTCweekly_R3.eps}
\includegraphics[width=0.49\textwidth]{figs/ETHweekly_R3.eps}
\caption{The superposed time series of log-returns $R_{\lambda_k}(j)$ calculated according to Eq.~(\ref{eigensignalweek}) for BTC and ETH. The U.S. macroeconomic reports related to the recurring patterns with increased volatility are marked.}
\label{fig::eigRweek}
\end{figure}

With the reference to the intraday dynamics portrayed in Figure~\ref{fig::eigR24}, it is discernible that the most recurrent structures emerge at 12:30 UTC. The intraweek decomposition, as shown in Figure~\ref{fig::eigRweek}, enables the identification of the specific days on which the amplified activity patterns appear. The superposed time series associated with the largest eigenvalue $\lambda_1$ suggests that the most pronounced synchronous activity happens on Fridays at 12:30 UTC, which correlates with the Nonfarm Payroll (NFP) report (see Table~\ref{tab::macro}). The second pattern associated with $\lambda_2$ in the case of ETH and $\lambda_3$ in the case of BTC is observed at 12:30 on Wednesdays and Thursdays when Consumer Price Index (CPI) and Producer Price Index (PPI) inflation reports are released. An additional pattern of the increased activity, corresponding to $\lambda_2$ for BTC and $\lambda_3$ for ETH, overlooked by the intraday decomposition, is manifest on Wednesdays at 18:00 UTC. This remains in agreement with the timing of the Federal Reserve statement releases.

The above findings indicate that the recurrent periods of the increased market activity coincide with the publication of the U.S. economic data. Their frequency can be a pivotal factor that influences the structure repeatability. Collectivity of the pattern related to the NFP reports is the strongest and attributable to its consistent monthly release schedule (Fridays at 12:30 UTC). In contrast, while the CPI and PPI reports are also published monthly, their release dates vary among the weekdays, excluding Mondays. The Federal Reserve meeting statements, associated here with $\lambda_2$ and $\lambda_3$, are disclosed approximately every 2.5 months, consistently on Wednesdays at 18:00 UTC.

\subsection{Correlations in individual years}

\begin{figure}
\includegraphics[width=0.49\textwidth]{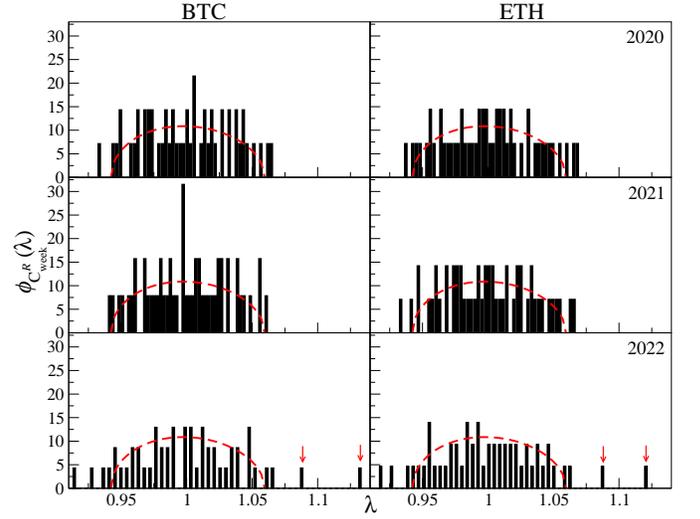}
\caption{Eigenvalue distribution $\phi^R_{\textrm{week}}$ obtained from the correlation matrix of intraday log-returns ${\bf C}^R_{\rm{day}}$ for BTC and ETH broken down by years. In each case, the Marchenko-Pastur distribution is shown (dashed red lines) and the eigenvalues outside it have been marked.}
\label{fig::MPdisryear}
\end{figure}

An interesting issue that emerges is related to the temporal uniformity of the identified activity patterns. The most important cross-correlated days were predominantly witnessed in 2022 (Table~\ref{tab::macro}). To ensure divergence of the patterns across each year, the intraweek log-returns were sequentially decomposed for each year individually. Figure~\ref{fig::MPdisryear} reveals that only in 2022 there were two eigenvalues that lay outside the random-matrix region. This means that during 2020 and 2021 the market volatility did not exhibit any recurrent intraweek pattern and that such patterns manifested themselves exclusively in the year 2022. Such disparate results could potentially be attributed to the significant correlations that emerged between the primary cryptocurrencies, in particular BTC and ETH, and the U.S. stock indices~\cite{WatorekM-2023a,Entropy2023}. Contrarily, WIN was identified as the most frequently traded cryptocurrency from the cohort of cryptocurrencies that were uncorrelated with the U.S. stock indices~\cite{Entropy2023}. It might elucidate the fact that all eigenvalues of the correlation matrix related to this cryptocurrency were found within the M-P region, which indicates the prevalence of the noisy intraday fluctuations and, thereby, the absence of any easily discernible repeatable patterns of the market dynamics. Furthermore, in the case of DOGE, the correlations with the U.S. indices were notably lower if compared to BTC and ETH~\cite{WatorekM-2023a}. This observation may explain the relatively reduced occurrence of repetitive patterns in the DOGE dynamics.

\begin{figure}
\includegraphics[width=0.49\textwidth]{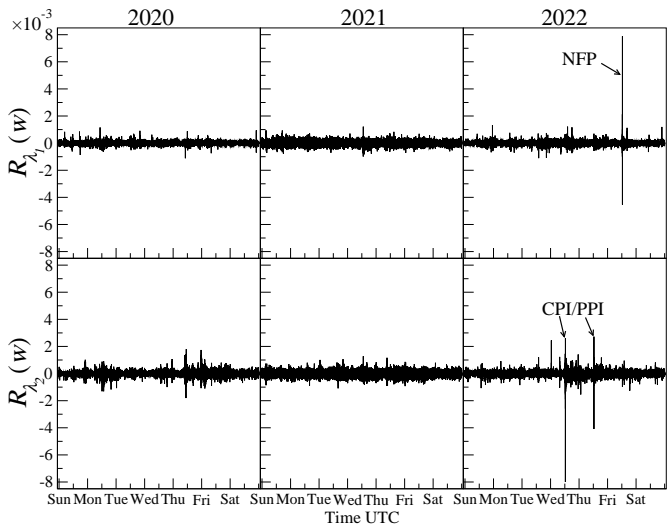}
\caption{The superposed time series of log-returns $R_{\lambda_k}(j)$ calculated according to Eq.~(\ref{eigensignalweek}) for BTC and ETH broken down by years. The U.S. macroeconomic reports related to the recurring patterns with increased volatility are marked.}
\label{fig::RVNweekly20-23}
\end{figure}

In line with the expectations derived from the number of eigenvalues situated outside the random matrix region for BTC in Fig.~\ref{fig::MPdisryear}, the superposed time series $R_{\lambda_k}(i)$, computed separately for each year from 2020 to 2022 display the absence of any recurrent pattern in 2020 and 2021 (Figure~\ref{fig::RVNweekly20-23}); they emerge only in 2022. The most pronounced collective effect associated with $\lambda_1$ is observed on Fridays at 12:30, a phenomenon that was previously attributed to the monthly releases of the NFP report. The patterns corresponding to $\lambda_2$ occur on Thursdays and Wednesdays at 12:30 UTC, which coincides with the inflation report releases. This is in agreement with the recent research findings that indicate a synchronized movement of BTC and ETH against the U.S. indices following the CPI announcements in 2022~\cite{WatorekM-2023a}. The inflation metrics were subject to an intense scrutiny in 2022, which induced substantial volatility across all financial markets~\cite{James2022inf}. While Bitcoin was fundamentally designed to serve as a safeguard against the policy of central banks and inflation owing to its a predetermined supply~\cite{Choi2022}, its observed response to the inflation data, akin to the stock indices, may suggest a deviation from such an assigned role. This can be interpreted as yet another indication that BTC has transformed into a volatile financial asset, which exhibits correlations with the other volatile assets.

\section{Summary}

By applying the correlation matrix formalism to decompose the cryptocurrency price dynamics into intraday and intraweek time series, this study demonstrates the efficacy of this formalism as a tool for identifying periods of increased market activity that are repeatable across different days or weeks. A comparison of the empirical eigenvalue distribution with the theoretical prediction for uncorrelated Gaussian time series expressed by the Marchenko-Pastur distribution allowed one here to discern the correlations pertaining to the synchronous behaviour of the cryptocurrency market from the overall noise. The price fluctuations (returns) of the most liquid cryptocurrencies -- BTC and ETH -- were identified to exhibit specific intraday and intraweek patterns. One such pattern involves the amplified market activity precisely at full hours and a couple of more subtle patterns involve similar effects at full quarters and half-hours. This is a cross-platform effect that cannot be confined to a unique trading platform. The algorithmic trading, a phenomenon well-recognized in the stock markets, offers a plausible explanation of this effect. It is worthwhile to note that, while the external factors considered here are responsible for the repeatable components of the market dynamics, the internal factors appear to be substantially random and unique as a good agreement between the empirical eigenvalue distributions in their bulk and the random matrix theory predictions expressed by the Marchenko-Pastur distribution.

Of a particular interest are the recurring patterns linked to the macroeconomic news releases. These moments are coupled with the increased activity of the market, which became evident predominantly in 2022. In the case of BTC and ETH, it could potentially be attributed to the emergence of the significant cross-correlation between the most liquid cryptocurrencies and the U.S. stock indices, in particular around the CPI report releases. It remains an open question whether these repeatable patterns observed in the cryptocurrency market are transmitted there through the cross-correlation with the U.S. indices, for which the amplified volatility is expected and well-documented, or they rather reflect an intrinsic reaction of the cryptocurrency market to such data. However, irrespective of the answer, which may form a basis for future research, the revealed dependence substantiates further that the cryptocurrency market has evolved into a valid and interrelated component of the global financial markets.

As a final remark, it needs to be emphasized that this methodology is largely universal and is expected to result in a similar decomposition of the dynamics for any market of reasonable liquidity.

\section*{Data Availability Statement}

The data are freely available on Binance exchange~\cite{Binance}.

\nocite{*}
\bibliography{manuscript_clean}

\begin{thebibliography}{88}%
\makeatletter
\providecommand \@ifxundefined [1]{%
 \@ifx{#1\undefined}
}%
\providecommand \@ifnum [1]{%
 \ifnum #1\expandafter \@firstoftwo
 \else \expandafter \@secondoftwo
 \fi
}%
\providecommand \@ifx [1]{%
 \ifx #1\expandafter \@firstoftwo
 \else \expandafter \@secondoftwo
 \fi
}%
\providecommand \natexlab [1]{#1}%
\providecommand \enquote  [1]{``#1''}%
\providecommand \bibnamefont  [1]{#1}%
\providecommand \bibfnamefont [1]{#1}%
\providecommand \citenamefont [1]{#1}%
\providecommand \href@noop [0]{\@secondoftwo}%
\providecommand \href [0]{\begingroup \@sanitize@url \@href}%
\providecommand \@href[1]{\@@startlink{#1}\@@href}%
\providecommand \@@href[1]{\endgroup#1\@@endlink}%
\providecommand \@sanitize@url [0]{\catcode `\\12\catcode `\$12\catcode
  `\&12\catcode `\#12\catcode `\^12\catcode `\_12\catcode `\%12\relax}%
\providecommand \@@startlink[1]{}%
\providecommand \@@endlink[0]{}%
\providecommand \url  [0]{\begingroup\@sanitize@url \@url }%
\providecommand \@url [1]{\endgroup\@href {#1}{\urlprefix }}%
\providecommand \urlprefix  [0]{URL }%
\providecommand \Eprint [0]{\href }%
\providecommand \doibase [0]{http://dx.doi.org/}%
\providecommand \selectlanguage [0]{\@gobble}%
\providecommand \bibinfo  [0]{\@secondoftwo}%
\providecommand \bibfield  [0]{\@secondoftwo}%
\providecommand \translation [1]{[#1]}%
\providecommand \BibitemOpen [0]{}%
\providecommand \bibitemStop [0]{}%
\providecommand \bibitemNoStop [0]{.\EOS\space}%
\providecommand \EOS [0]{\spacefactor3000\relax}%
\providecommand \BibitemShut  [1]{\csname bibitem#1\endcsname}%
\let\auto@bib@innerbib\@empty
\bibitem [{\citenamefont {Gerety}\ and\ \citenamefont
  {Mulherin}(1991)}]{Gerety1991}%
  \BibitemOpen
  \bibfield  {author} {\bibinfo {author} {\bibfnamefont {M.~S.}\ \bibnamefont
  {Gerety}}\ and\ \bibinfo {author} {\bibfnamefont {J.~H.}\ \bibnamefont
  {Mulherin}},\ }\bibfield  {title} {\enquote {\bibinfo {title} {Patterns in
  intraday stock market volatility, past and present},}\ }\href@noop {}
  {\bibfield  {journal} {\bibinfo  {journal} {Financial Analysts Journal}\
  }\textbf {\bibinfo {volume} {47}},\ \bibinfo {pages} {71--79} (\bibinfo
  {year} {1991})}\BibitemShut {NoStop}%
\bibitem [{\citenamefont {Andersen}\ and\ \citenamefont
  {Bollerslev}(1998)}]{Andersen1998}%
  \BibitemOpen
  \bibfield  {author} {\bibinfo {author} {\bibfnamefont {T.~G.}\ \bibnamefont
  {Andersen}}\ and\ \bibinfo {author} {\bibfnamefont {T.}~\bibnamefont
  {Bollerslev}},\ }\bibfield  {title} {\enquote {\bibinfo {title} {Deutsche
  mark–dollar volatility: Intraday activity patterns, macroeconomic
  announcements, and longer run dependencies},}\ }\href@noop {} {\bibfield
  {journal} {\bibinfo  {journal} {The Journal of Finance}\ }\textbf {\bibinfo
  {volume} {53}},\ \bibinfo {pages} {219--265} (\bibinfo {year}
  {1998})}\BibitemShut {NoStop}%
\bibitem [{\citenamefont {Bollerslev}, \citenamefont {Cai},\ and\ \citenamefont
  {Song}(2000)}]{Bollerslev2000}%
  \BibitemOpen
  \bibfield  {author} {\bibinfo {author} {\bibfnamefont {T.}~\bibnamefont
  {Bollerslev}}, \bibinfo {author} {\bibfnamefont {J.}~\bibnamefont {Cai}}, \
  and\ \bibinfo {author} {\bibfnamefont {F.~M.}\ \bibnamefont {Song}},\
  }\bibfield  {title} {\enquote {\bibinfo {title} {Intraday periodicity, long
  memory volatility, and macroeconomic announcement effects in the us treasury
  bond market},}\ }\href@noop {} {\bibfield  {journal} {\bibinfo  {journal}
  {Journal of Empirical Finance}\ }\textbf {\bibinfo {volume} {7}},\ \bibinfo
  {pages} {37--55} (\bibinfo {year} {2000})}\BibitemShut {NoStop}%
\bibitem [{\citenamefont {Andersen}\ \emph {et~al.}(2003)\citenamefont
  {Andersen}, \citenamefont {Bollerslev}, \citenamefont {Diebold},\ and\
  \citenamefont {Vega}}]{Andersen2003}%
  \BibitemOpen
  \bibfield  {author} {\bibinfo {author} {\bibfnamefont {T.~G.}\ \bibnamefont
  {Andersen}}, \bibinfo {author} {\bibfnamefont {T.}~\bibnamefont
  {Bollerslev}}, \bibinfo {author} {\bibfnamefont {F.~X.}\ \bibnamefont
  {Diebold}}, \ and\ \bibinfo {author} {\bibfnamefont {C.}~\bibnamefont
  {Vega}},\ }\bibfield  {title} {\enquote {\bibinfo {title} {Micro effects of
  macro announcements: Real-time price discovery in foreign exchange},}\
  }\href@noop {} {\bibfield  {journal} {\bibinfo  {journal} {American Economic
  Review}\ }\textbf {\bibinfo {volume} {93}},\ \bibinfo {pages} {38--62}
  (\bibinfo {year} {2003})}\BibitemShut {NoStop}%
\bibitem [{\citenamefont {Evans}\ and\ \citenamefont
  {Speight}(2010)}]{Evans2010}%
  \BibitemOpen
  \bibfield  {author} {\bibinfo {author} {\bibfnamefont {K.}~\bibnamefont
  {Evans}}\ and\ \bibinfo {author} {\bibfnamefont {A.}~\bibnamefont
  {Speight}},\ }\bibfield  {title} {\enquote {\bibinfo {title} {International
  macroeconomic announcements and intraday euro exchange rate volatility},}\
  }\href@noop {} {\bibfield  {journal} {\bibinfo  {journal} {Journal of the
  Japanese and International Economies}\ }\textbf {\bibinfo {volume} {24}},\
  \bibinfo {pages} {552--568} (\bibinfo {year} {2010})}\BibitemShut {NoStop}%
\bibitem [{\citenamefont {Evans}(2011)}]{Evans2011}%
  \BibitemOpen
  \bibfield  {author} {\bibinfo {author} {\bibfnamefont {K.~P.}\ \bibnamefont
  {Evans}},\ }\bibfield  {title} {\enquote {\bibinfo {title} {Intraday jumps
  and us macroeconomic news announcements},}\ }\href@noop {} {\bibfield
  {journal} {\bibinfo  {journal} {Journal of Banking \& Finance}\ }\textbf
  {\bibinfo {volume} {35}},\ \bibinfo {pages} {2511--2527} (\bibinfo {year}
  {2011})}\BibitemShut {NoStop}%
\bibitem [{\citenamefont {Harju}\ and\ \citenamefont
  {Hussain}(2011)}]{Kari2011}%
  \BibitemOpen
  \bibfield  {author} {\bibinfo {author} {\bibfnamefont {K.}~\bibnamefont
  {Harju}}\ and\ \bibinfo {author} {\bibfnamefont {S.~M.}\ \bibnamefont
  {Hussain}},\ }\bibfield  {title} {\enquote {\bibinfo {title} {Intraday
  seasonalities and macroeconomic news announcements},}\ }\href@noop {}
  {\bibfield  {journal} {\bibinfo  {journal} {European Financial Management}\
  }\textbf {\bibinfo {volume} {17}},\ \bibinfo {pages} {367--390} (\bibinfo
  {year} {2011})}\BibitemShut {NoStop}%
\bibitem [{\citenamefont {Hussain}(2011)}]{HUSSAIN2011}%
  \BibitemOpen
  \bibfield  {author} {\bibinfo {author} {\bibfnamefont {S.~M.}\ \bibnamefont
  {Hussain}},\ }\bibfield  {title} {\enquote {\bibinfo {title} {Simultaneous
  monetary policy announcements and international stock markets response: An
  intraday analysis},}\ }\href@noop {} {\bibfield  {journal} {\bibinfo
  {journal} {Journal of Banking \& Finance}\ }\textbf {\bibinfo {volume}
  {35}},\ \bibinfo {pages} {752--764} (\bibinfo {year} {2011})}\BibitemShut
  {NoStop}%
\bibitem [{\citenamefont {Kenourgios}, \citenamefont {Papadamou},\ and\
  \citenamefont {Dimitriou}(2015)}]{Kenourgios2015}%
  \BibitemOpen
  \bibfield  {author} {\bibinfo {author} {\bibfnamefont {D.}~\bibnamefont
  {Kenourgios}}, \bibinfo {author} {\bibfnamefont {S.}~\bibnamefont
  {Papadamou}}, \ and\ \bibinfo {author} {\bibfnamefont {D.}~\bibnamefont
  {Dimitriou}},\ }\bibfield  {title} {\enquote {\bibinfo {title} {Intraday
  exchange rate volatility transmissions across {QE} announcements},}\
  }\href@noop {} {\bibfield  {journal} {\bibinfo  {journal} {Finance Research
  Letters}\ }\textbf {\bibinfo {volume} {14}},\ \bibinfo {pages} {128--134}
  (\bibinfo {year} {2015})}\BibitemShut {NoStop}%
\bibitem [{\citenamefont {G\k{e}barowski}\ \emph {et~al.}(2019)\citenamefont
  {G\k{e}barowski}, \citenamefont {O\'swi\k{e}cimka}, \citenamefont
  {W\k{a}torek},\ and\ \citenamefont {Dro\.{z}d\.{z}}}]{gebarowski2019}%
  \BibitemOpen
  \bibfield  {author} {\bibinfo {author} {\bibfnamefont {R.}~\bibnamefont
  {G\k{e}barowski}}, \bibinfo {author} {\bibfnamefont {P.}~\bibnamefont
  {O\'swi\k{e}cimka}}, \bibinfo {author} {\bibfnamefont {M.}~\bibnamefont
  {W\k{a}torek}}, \ and\ \bibinfo {author} {\bibfnamefont {S.}~\bibnamefont
  {Dro\.{z}d\.{z}}},\ }\bibfield  {title} {\enquote {\bibinfo {title}
  {Detecting correlations and triangular arbitrage opportunities in the {Forex}
  by means of multifractal detrended cross-correlations analysis},}\
  }\href@noop {} {\bibfield  {journal} {\bibinfo  {journal} {Nonlinear
  Dynamics}\ }\textbf {\bibinfo {volume} {98}},\ \bibinfo {pages} {2349--2364}
  (\bibinfo {year} {2019})}\BibitemShut {NoStop}%
\bibitem [{\citenamefont {Gjerstad}\ \emph {et~al.}(2021)\citenamefont
  {Gjerstad}, \citenamefont {Meyn}, \citenamefont {Molnár},\ and\
  \citenamefont {Næss}}]{Gjerstad2021}%
  \BibitemOpen
  \bibfield  {author} {\bibinfo {author} {\bibfnamefont {P.}~\bibnamefont
  {Gjerstad}}, \bibinfo {author} {\bibfnamefont {P.~F.}\ \bibnamefont {Meyn}},
  \bibinfo {author} {\bibfnamefont {P.}~\bibnamefont {Molnár}}, \ and\
  \bibinfo {author} {\bibfnamefont {T.~D.}\ \bibnamefont {Næss}},\ }\bibfield
  {title} {\enquote {\bibinfo {title} {Do {President Trump's} tweets affect
  financial markets?}}\ }\href@noop {} {\bibfield  {journal} {\bibinfo
  {journal} {Decision Support Systems}\ }\textbf {\bibinfo {volume} {147}},\
  \bibinfo {pages} {113577} (\bibinfo {year} {2021})}\BibitemShut {NoStop}%
\bibitem [{\citenamefont {Andersen}, \citenamefont {Bollerslev},\ and\
  \citenamefont {Cai}(2000)}]{Andersen2000}%
  \BibitemOpen
  \bibfield  {author} {\bibinfo {author} {\bibfnamefont {T.~G.}\ \bibnamefont
  {Andersen}}, \bibinfo {author} {\bibfnamefont {T.}~\bibnamefont
  {Bollerslev}}, \ and\ \bibinfo {author} {\bibfnamefont {J.}~\bibnamefont
  {Cai}},\ }\bibfield  {title} {\enquote {\bibinfo {title} {Intraday and
  interday volatility in the {Japanese} stock market},}\ }\href@noop {}
  {\bibfield  {journal} {\bibinfo  {journal} {Journal of International
  Financial Markets, Institutions and Money}\ }\textbf {\bibinfo {volume}
  {10}},\ \bibinfo {pages} {107--130} (\bibinfo {year} {2000})}\BibitemShut
  {NoStop}%
\bibitem [{\citenamefont {Gubiec}\ and\ \citenamefont
  {Wiliński}(2015)}]{Gubiec2015}%
  \BibitemOpen
  \bibfield  {author} {\bibinfo {author} {\bibfnamefont {T.}~\bibnamefont
  {Gubiec}}\ and\ \bibinfo {author} {\bibfnamefont {M.}~\bibnamefont
  {Wiliński}},\ }\bibfield  {title} {\enquote {\bibinfo {title} {Intra-day
  variability of the stock market activity versus stationarity of the financial
  time series},}\ }\href@noop {} {\bibfield  {journal} {\bibinfo  {journal}
  {Physica A}\ }\textbf {\bibinfo {volume} {432}},\ \bibinfo {pages} {216--221}
  (\bibinfo {year} {2015})}\BibitemShut {NoStop}%
\bibitem [{\citenamefont {Andersen}, \citenamefont {Thyrsgaard},\ and\
  \citenamefont {Todorov}(2019)}]{Andersen2019}%
  \BibitemOpen
  \bibfield  {author} {\bibinfo {author} {\bibfnamefont {T.~G.}\ \bibnamefont
  {Andersen}}, \bibinfo {author} {\bibfnamefont {M.}~\bibnamefont
  {Thyrsgaard}}, \ and\ \bibinfo {author} {\bibfnamefont {V.}~\bibnamefont
  {Todorov}},\ }\bibfield  {title} {\enquote {\bibinfo {title} {Time-varying
  periodicity in intraday volatility},}\ }\href@noop {} {\bibfield  {journal}
  {\bibinfo  {journal} {Journal of the American Statistical Association}\
  }\textbf {\bibinfo {volume} {114}},\ \bibinfo {pages} {1695--1707} (\bibinfo
  {year} {2019})}\BibitemShut {NoStop}%
\bibitem [{\citenamefont {Berument}\ and\ \citenamefont
  {Kiymaz}(2001)}]{berument2001day}%
  \BibitemOpen
  \bibfield  {author} {\bibinfo {author} {\bibfnamefont {H.}~\bibnamefont
  {Berument}}\ and\ \bibinfo {author} {\bibfnamefont {H.}~\bibnamefont
  {Kiymaz}},\ }\bibfield  {title} {\enquote {\bibinfo {title} {The day of the
  week effect on stock market volatility},}\ }\href@noop {} {\bibfield
  {journal} {\bibinfo  {journal} {Journal of Economics and Finance}\ }\textbf
  {\bibinfo {volume} {25}},\ \bibinfo {pages} {181--193} (\bibinfo {year}
  {2001})}\BibitemShut {NoStop}%
\bibitem [{\citenamefont {Dacorogna}\ \emph {et~al.}(1993)\citenamefont
  {Dacorogna}, \citenamefont {Müller}, \citenamefont {Nagler}, \citenamefont
  {Olsen},\ and\ \citenamefont {Pictet}}]{Dacorogna1993}%
  \BibitemOpen
  \bibfield  {author} {\bibinfo {author} {\bibfnamefont {M.~M.}\ \bibnamefont
  {Dacorogna}}, \bibinfo {author} {\bibfnamefont {U.~A.}\ \bibnamefont
  {Müller}}, \bibinfo {author} {\bibfnamefont {R.~J.}\ \bibnamefont {Nagler}},
  \bibinfo {author} {\bibfnamefont {R.~B.}\ \bibnamefont {Olsen}}, \ and\
  \bibinfo {author} {\bibfnamefont {O.~V.}\ \bibnamefont {Pictet}},\ }\bibfield
   {title} {\enquote {\bibinfo {title} {A geographical model for the daily and
  weekly seasonal volatility in the foreign exchange market},}\ }\href@noop {}
  {\bibfield  {journal} {\bibinfo  {journal} {Journal of International Money
  and Finance}\ }\textbf {\bibinfo {volume} {12}},\ \bibinfo {pages} {413--438}
  (\bibinfo {year} {1993})}\BibitemShut {NoStop}%
\bibitem [{\citenamefont {Zhang}(2018)}]{Zhang2018}%
  \BibitemOpen
  \bibfield  {author} {\bibinfo {author} {\bibfnamefont {H.}~\bibnamefont
  {Zhang}},\ }\bibfield  {title} {\enquote {\bibinfo {title} {Intraday patterns
  in foreign exchange returns and realized volatility},}\ }\href@noop {}
  {\bibfield  {journal} {\bibinfo  {journal} {Finance Research Letters}\
  }\textbf {\bibinfo {volume} {27}},\ \bibinfo {pages} {99--104} (\bibinfo
  {year} {2018})}\BibitemShut {NoStop}%
\bibitem [{\citenamefont {Bu}(2014)}]{Bu2014}%
  \BibitemOpen
  \bibfield  {author} {\bibinfo {author} {\bibfnamefont {H.}~\bibnamefont
  {Bu}},\ }\bibfield  {title} {\enquote {\bibinfo {title} {Effect of inventory
  announcements on crude oil price volatility},}\ }\href@noop {} {\bibfield
  {journal} {\bibinfo  {journal} {Energy Economics}\ }\textbf {\bibinfo
  {volume} {46}},\ \bibinfo {pages} {485--494} (\bibinfo {year}
  {2014})}\BibitemShut {NoStop}%
\bibitem [{\citenamefont {Alves}\ \emph {et~al.}(2020)\citenamefont {Alves},
  \citenamefont {Sigaki}, \citenamefont {Perc},\ and\ \citenamefont
  {Ribeiro}}]{alves2020}%
  \BibitemOpen
  \bibfield  {author} {\bibinfo {author} {\bibfnamefont {L.~G.}\ \bibnamefont
  {Alves}}, \bibinfo {author} {\bibfnamefont {H.~Y.}\ \bibnamefont {Sigaki}},
  \bibinfo {author} {\bibfnamefont {M.}~\bibnamefont {Perc}}, \ and\ \bibinfo
  {author} {\bibfnamefont {H.~V.}\ \bibnamefont {Ribeiro}},\ }\bibfield
  {title} {\enquote {\bibinfo {title} {Collective dynamics of stock market
  efficiency},}\ }\href@noop {} {\bibfield  {journal} {\bibinfo  {journal}
  {Scientific Reports}\ }\textbf {\bibinfo {volume} {10}},\ \bibinfo {pages}
  {21992} (\bibinfo {year} {2020})}\BibitemShut {NoStop}%
\bibitem [{CoinMarketCap()}]{coinmarket}%
  \BibitemOpen
  CoinMarketCap,\ \href@noop {} {\enquote {\bibinfo {title}
  {{CoinMarketCap}},}\ }\bibinfo {howpublished}
  {\url{https://coinmarketcap.com}}\BibitemShut {NoStop}%
\bibitem [{\citenamefont {Gerlach}, \citenamefont {Demos},\ and\ \citenamefont
  {Sornette}(2019)}]{Gerlach2018}%
  \BibitemOpen
  \bibfield  {author} {\bibinfo {author} {\bibfnamefont {J.-C.}\ \bibnamefont
  {Gerlach}}, \bibinfo {author} {\bibfnamefont {G.}~\bibnamefont {Demos}}, \
  and\ \bibinfo {author} {\bibfnamefont {D.}~\bibnamefont {Sornette}},\
  }\bibfield  {title} {\enquote {\bibinfo {title} {Dissection of {Bitcoin's}
  multiscale bubble history from {J}anuary 2012 to {F}ebruary 2018},}\
  }\href@noop {} {\bibfield  {journal} {\bibinfo  {journal} {Royal Society Open
  Science}\ }\textbf {\bibinfo {volume} {6}},\ \bibinfo {pages} {180643}
  (\bibinfo {year} {2019})}\BibitemShut {NoStop}%
\bibitem [{\citenamefont {Bellon}\ and\ \citenamefont
  {Figuerola-Ferretti}(2022)}]{Bellon2022}%
  \BibitemOpen
  \bibfield  {author} {\bibinfo {author} {\bibfnamefont {C.}~\bibnamefont
  {Bellon}}\ and\ \bibinfo {author} {\bibfnamefont {I.}~\bibnamefont
  {Figuerola-Ferretti}},\ }\bibfield  {title} {\enquote {\bibinfo {title}
  {Bubbles in {Ethereum}},}\ }\href@noop {} {\bibfield  {journal} {\bibinfo
  {journal} {Finance Research Letters}\ }\textbf {\bibinfo {volume} {46}},\
  \bibinfo {pages} {102387} (\bibinfo {year} {2022})}\BibitemShut {NoStop}%
\bibitem [{\citenamefont {Baz\'{a}n-Palomino}(2022)}]{Palomino2022}%
  \BibitemOpen
  \bibfield  {author} {\bibinfo {author} {\bibfnamefont {W.}~\bibnamefont
  {Baz\'{a}n-Palomino}},\ }\bibfield  {title} {\enquote {\bibinfo {title}
  {Interdependence, contagion and speculative bubbles in cryptocurrency
  markets},}\ }\href@noop {} {\bibfield  {journal} {\bibinfo  {journal}
  {Finance Research Letters}\ }\textbf {\bibinfo {volume} {49}},\ \bibinfo
  {pages} {103132} (\bibinfo {year} {2022})}\BibitemShut {NoStop}%
\bibitem [{\citenamefont {Wang}\ \emph {et~al.}(2022)\citenamefont {Wang},
  \citenamefont {Horky}, \citenamefont {Baals}, \citenamefont {Lucey},\ and\
  \citenamefont {Vigne}}]{Wang2022}%
  \BibitemOpen
  \bibfield  {author} {\bibinfo {author} {\bibfnamefont {Y.}~\bibnamefont
  {Wang}}, \bibinfo {author} {\bibfnamefont {F.}~\bibnamefont {Horky}},
  \bibinfo {author} {\bibfnamefont {L.~J.}\ \bibnamefont {Baals}}, \bibinfo
  {author} {\bibfnamefont {B.~M.}\ \bibnamefont {Lucey}}, \ and\ \bibinfo
  {author} {\bibfnamefont {S.~A.}\ \bibnamefont {Vigne}},\ }\bibfield  {title}
  {\enquote {\bibinfo {title} {Bubbles all the way down? {Detecting} and
  date-stamping bubble behaviours in {NFT} and {DeFi} markets},}\ }\href@noop
  {} {\bibfield  {journal} {\bibinfo  {journal} {Journal of Chinese Economic
  and Business Studies}\ }\textbf {\bibinfo {volume} {20}},\ \bibinfo {pages}
  {415--436} (\bibinfo {year} {2022})}\BibitemShut {NoStop}%
\bibitem [{\citenamefont {Charoenwong}\ and\ \citenamefont
  {Bernardi}(2021)}]{charoenwong2021decade}%
  \BibitemOpen
  \bibfield  {author} {\bibinfo {author} {\bibfnamefont {B.}~\bibnamefont
  {Charoenwong}}\ and\ \bibinfo {author} {\bibfnamefont {M.}~\bibnamefont
  {Bernardi}},\ }\bibfield  {title} {\enquote {\bibinfo {title} {A decade of
  cryptocurrency ‘hacks’: 2011--2021},}\ }\href@noop {} {\bibfield
  {journal} {\bibinfo  {journal} {Available at SSRN 3944435}\ } (\bibinfo
  {year} {2021})}\BibitemShut {NoStop}%
\bibitem [{\citenamefont {Fu}\ \emph {et~al.}(2022)\citenamefont {Fu},
  \citenamefont {Wang}, \citenamefont {Yu},\ and\ \citenamefont
  {Chen}}]{fu2022ftx}%
  \BibitemOpen
  \bibfield  {author} {\bibinfo {author} {\bibfnamefont {S.}~\bibnamefont
  {Fu}}, \bibinfo {author} {\bibfnamefont {Q.}~\bibnamefont {Wang}}, \bibinfo
  {author} {\bibfnamefont {J.}~\bibnamefont {Yu}}, \ and\ \bibinfo {author}
  {\bibfnamefont {S.}~\bibnamefont {Chen}},\ }\href@noop {} {\enquote {\bibinfo
  {title} {{FTX} collapse: A {Ponzi} story},}\ } (\bibinfo {year} {2022}),\
  \Eprint {http://arxiv.org/abs/2212.09436} {arXiv:2212.09436 [cs.CR]}
  \BibitemShut {NoStop}%
\bibitem [{\citenamefont {Briola}\ \emph {et~al.}(2023)\citenamefont {Briola},
  \citenamefont {Vidal-Tomas}, \citenamefont {Wang},\ and\ \citenamefont
  {Aste}}]{Briola2023}%
  \BibitemOpen
  \bibfield  {author} {\bibinfo {author} {\bibfnamefont {A.}~\bibnamefont
  {Briola}}, \bibinfo {author} {\bibfnamefont {D.}~\bibnamefont {Vidal-Tomas}},
  \bibinfo {author} {\bibfnamefont {Y.}~\bibnamefont {Wang}}, \ and\ \bibinfo
  {author} {\bibfnamefont {T.}~\bibnamefont {Aste}},\ }\bibfield  {title}
  {\enquote {\bibinfo {title} {Anatomy of a stablecoin’s failure: {The
  Terra-Luna} case},}\ }\href@noop {} {\bibfield  {journal} {\bibinfo
  {journal} {Finance Research Letters}\ }\textbf {\bibinfo {volume} {51}},\
  \bibinfo {pages} {103358} (\bibinfo {year} {2023})}\BibitemShut {NoStop}%
\bibitem [{\citenamefont {Vidal-Tomás}, \citenamefont {Briola},\ and\
  \citenamefont {Aste}(2023)}]{vidaltomas2023ftxs}%
  \BibitemOpen
  \bibfield  {author} {\bibinfo {author} {\bibfnamefont {D.}~\bibnamefont
  {Vidal-Tomás}}, \bibinfo {author} {\bibfnamefont {A.}~\bibnamefont
  {Briola}}, \ and\ \bibinfo {author} {\bibfnamefont {T.}~\bibnamefont
  {Aste}},\ }\bibfield  {title} {\enquote {\bibinfo {title} {{FTX’s} downfall
  and {Binance’s} consolidation: {The} fragility of centralised digital
  finance},}\ }\href@noop {} {\bibfield  {journal} {\bibinfo  {journal}
  {Physica A}\ }\textbf {\bibinfo {volume} {625}},\ \bibinfo {pages} {129044}
  (\bibinfo {year} {2023})}\BibitemShut {NoStop}%
\bibitem [{\citenamefont {Bariviera}, \citenamefont {Zunino},\ and\
  \citenamefont {Rosso}(2018)}]{bariviera2018analysis}%
  \BibitemOpen
  \bibfield  {author} {\bibinfo {author} {\bibfnamefont {A.~F.}\ \bibnamefont
  {Bariviera}}, \bibinfo {author} {\bibfnamefont {L.}~\bibnamefont {Zunino}}, \
  and\ \bibinfo {author} {\bibfnamefont {O.~A.}\ \bibnamefont {Rosso}},\
  }\bibfield  {title} {\enquote {\bibinfo {title} {An analysis of
  high-frequency cryptocurrencies prices dynamics using
  permutation-information-theory quantifiers},}\ }\href@noop {} {\bibfield
  {journal} {\bibinfo  {journal} {Chaos}\ }\textbf {\bibinfo {volume} {28}},\
  \bibinfo {pages} {075511} (\bibinfo {year} {2018})}\BibitemShut {NoStop}%
\bibitem [{\citenamefont {Dro\.{z}d\.{z}}\ \emph {et~al.}(2018)\citenamefont
  {Dro\.{z}d\.{z}}, \citenamefont {G\k{e}barowski}, \citenamefont {Minati},
  \citenamefont {O\'swi\k{e}cimka},\ and\ \citenamefont
  {W\k{a}torek}}]{DrozdzBTC2018}%
  \BibitemOpen
  \bibfield  {author} {\bibinfo {author} {\bibfnamefont {S.}~\bibnamefont
  {Dro\.{z}d\.{z}}}, \bibinfo {author} {\bibfnamefont {R.}~\bibnamefont
  {G\k{e}barowski}}, \bibinfo {author} {\bibfnamefont {L.}~\bibnamefont
  {Minati}}, \bibinfo {author} {\bibfnamefont {P.}~\bibnamefont
  {O\'swi\k{e}cimka}}, \ and\ \bibinfo {author} {\bibfnamefont
  {M.}~\bibnamefont {W\k{a}torek}},\ }\bibfield  {title} {\enquote {\bibinfo
  {title} {{Bitcoin market route to maturity? Evidence from return
  fluctuations, temporal correlations and multiscaling effects}},}\ }\href@noop
  {} {\bibfield  {journal} {\bibinfo  {journal} {Chaos}\ }\textbf {\bibinfo
  {volume} {28}},\ \bibinfo {pages} {071101} (\bibinfo {year}
  {2018})}\BibitemShut {NoStop}%
\bibitem [{\citenamefont {Sigaki}, \citenamefont {Perc},\ and\ \citenamefont
  {Ribeiro}(2019)}]{sigaki2019}%
  \BibitemOpen
  \bibfield  {author} {\bibinfo {author} {\bibfnamefont {H.~Y.}\ \bibnamefont
  {Sigaki}}, \bibinfo {author} {\bibfnamefont {M.}~\bibnamefont {Perc}}, \ and\
  \bibinfo {author} {\bibfnamefont {H.~V.}\ \bibnamefont {Ribeiro}},\
  }\bibfield  {title} {\enquote {\bibinfo {title} {Clustering patterns in
  efficiency and the coming-of-age of the cryptocurrency market},}\ }\href@noop
  {} {\bibfield  {journal} {\bibinfo  {journal} {Scientific Reports}\ }\textbf
  {\bibinfo {volume} {9}},\ \bibinfo {pages} {1440} (\bibinfo {year}
  {2019})}\BibitemShut {NoStop}%
\bibitem [{\citenamefont {W\k{a}torek}\ \emph {et~al.}(2021)\citenamefont
  {W\k{a}torek}, \citenamefont {Dro\.{z}d\.{z}}, \citenamefont {Kwapie\'n},
  \citenamefont {Minati}, \citenamefont {O\'swi\k{e}cimka},\ and\ \citenamefont
  {Stanuszek}}]{watorek2021}%
  \BibitemOpen
  \bibfield  {author} {\bibinfo {author} {\bibfnamefont {M.}~\bibnamefont
  {W\k{a}torek}}, \bibinfo {author} {\bibfnamefont {S.}~\bibnamefont
  {Dro\.{z}d\.{z}}}, \bibinfo {author} {\bibfnamefont {J.}~\bibnamefont
  {Kwapie\'n}}, \bibinfo {author} {\bibfnamefont {L.}~\bibnamefont {Minati}},
  \bibinfo {author} {\bibfnamefont {P.}~\bibnamefont {O\'swi\k{e}cimka}}, \
  and\ \bibinfo {author} {\bibfnamefont {M.}~\bibnamefont {Stanuszek}},\
  }\bibfield  {title} {\enquote {\bibinfo {title} {Multiscale characteristics
  of the emerging global cryptocurrency market},}\ }\href@noop {} {\bibfield
  {journal} {\bibinfo  {journal} {Physics Reports}\ }\textbf {\bibinfo {volume}
  {901}},\ \bibinfo {pages} {1--82} (\bibinfo {year} {2021})}\BibitemShut
  {NoStop}%
\bibitem [{\citenamefont {Kwapie{\'n}}\ \emph {et~al.}(2022)\citenamefont
  {Kwapie{\'n}}, \citenamefont {W{\k{a}}torek}, \citenamefont {Bezbradica},
  \citenamefont {Crane}, \citenamefont {Tan~Mai},\ and\ \citenamefont
  {Dro{\.z}d{\.z}}}]{KwapienJ-2022a}%
  \BibitemOpen
  \bibfield  {author} {\bibinfo {author} {\bibfnamefont {J.}~\bibnamefont
  {Kwapie{\'n}}}, \bibinfo {author} {\bibfnamefont {M.}~\bibnamefont
  {W{\k{a}}torek}}, \bibinfo {author} {\bibfnamefont {M.}~\bibnamefont
  {Bezbradica}}, \bibinfo {author} {\bibfnamefont {M.}~\bibnamefont {Crane}},
  \bibinfo {author} {\bibfnamefont {T.}~\bibnamefont {Tan~Mai}}, \ and\
  \bibinfo {author} {\bibfnamefont {S.}~\bibnamefont {Dro{\.z}d{\.z}}},\
  }\bibfield  {title} {\enquote {\bibinfo {title} {Analysis of
  inter-transaction time fluctuations in the cryptocurrency market},}\
  }\href@noop {} {\bibfield  {journal} {\bibinfo  {journal} {Chaos}\ }\textbf
  {\bibinfo {volume} {32}},\ \bibinfo {pages} {083142} (\bibinfo {year}
  {2022})}\BibitemShut {NoStop}%
\bibitem [{\citenamefont {Garcin}(2023)}]{garcin2023complexity}%
  \BibitemOpen
  \bibfield  {author} {\bibinfo {author} {\bibfnamefont {M.}~\bibnamefont
  {Garcin}},\ }\href@noop {} {\enquote {\bibinfo {title} {Complexity measure,
  kernel density estimation, bandwidth selection, and the efficient market
  hypothesis},}\ } (\bibinfo {year} {2023}),\ \Eprint
  {http://arxiv.org/abs/2305.13123} {arXiv:2305.13123 [q-fin.ST]} \BibitemShut
  {NoStop}%
\bibitem [{\citenamefont {Pessa}, \citenamefont {Perc},\ and\ \citenamefont
  {Ribeiro}(2023)}]{pessa2023}%
  \BibitemOpen
  \bibfield  {author} {\bibinfo {author} {\bibfnamefont {A.~A.}\ \bibnamefont
  {Pessa}}, \bibinfo {author} {\bibfnamefont {M.}~\bibnamefont {Perc}}, \ and\
  \bibinfo {author} {\bibfnamefont {H.~V.}\ \bibnamefont {Ribeiro}},\
  }\bibfield  {title} {\enquote {\bibinfo {title} {Age and market
  capitalization drive large price variations of cryptocurrencies},}\
  }\href@noop {} {\bibfield  {journal} {\bibinfo  {journal} {Scientific
  Reports}\ }\textbf {\bibinfo {volume} {13}},\ \bibinfo {pages} {3351}
  (\bibinfo {year} {2023})}\BibitemShut {NoStop}%
\bibitem [{\citenamefont {Manavi}\ \emph {et~al.}(2020)\citenamefont {Manavi},
  \citenamefont {Jafari}, \citenamefont {Rouhani},\ and\ \citenamefont
  {Ausloos}}]{Manavi2020}%
  \BibitemOpen
  \bibfield  {author} {\bibinfo {author} {\bibfnamefont {S.~A.}\ \bibnamefont
  {Manavi}}, \bibinfo {author} {\bibfnamefont {G.}~\bibnamefont {Jafari}},
  \bibinfo {author} {\bibfnamefont {S.}~\bibnamefont {Rouhani}}, \ and\
  \bibinfo {author} {\bibfnamefont {M.}~\bibnamefont {Ausloos}},\ }\bibfield
  {title} {\enquote {\bibinfo {title} {Demythifying the belief in
  cryptocurrencies decentralized aspects. {A} study of cryptocurrencies time
  cross-correlations with common currencies, commodities and financial
  indices},}\ }\href@noop {} {\bibfield  {journal} {\bibinfo  {journal}
  {Physica A}\ }\textbf {\bibinfo {volume} {556}},\ \bibinfo {pages} {124759}
  (\bibinfo {year} {2020})}\BibitemShut {NoStop}%
\bibitem [{\citenamefont {Balcilar}, \citenamefont {Ozdemir},\ and\
  \citenamefont {Agan}(2022)}]{Balcilar2022}%
  \BibitemOpen
  \bibfield  {author} {\bibinfo {author} {\bibfnamefont {M.}~\bibnamefont
  {Balcilar}}, \bibinfo {author} {\bibfnamefont {H.}~\bibnamefont {Ozdemir}}, \
  and\ \bibinfo {author} {\bibfnamefont {B.}~\bibnamefont {Agan}},\ }\bibfield
  {title} {\enquote {\bibinfo {title} {Effects of {COVID-19} on cryptocurrency
  and emerging market connectedness: {E}mpirical evidence from quantile,
  frequency, and lasso networks},}\ }\href@noop {} {\bibfield  {journal}
  {\bibinfo  {journal} {Physica A}\ }\textbf {\bibinfo {volume} {604}},\
  \bibinfo {pages} {127885} (\bibinfo {year} {2022})}\BibitemShut {NoStop}%
\bibitem [{\citenamefont {Wang}, \citenamefont {Liu},\ and\ \citenamefont
  {Wu}(2022)}]{Wnag2022Entr}%
  \BibitemOpen
  \bibfield  {author} {\bibinfo {author} {\bibfnamefont {P.}~\bibnamefont
  {Wang}}, \bibinfo {author} {\bibfnamefont {X.}~\bibnamefont {Liu}}, \ and\
  \bibinfo {author} {\bibfnamefont {S.}~\bibnamefont {Wu}},\ }\bibfield
  {title} {\enquote {\bibinfo {title} {Dynamic linkage between {Bitcoin} and
  traditional financial assets: {A} comparative analysis of different time
  frequencies},}\ }\href@noop {} {\bibfield  {journal} {\bibinfo  {journal}
  {Entropy}\ }\textbf {\bibinfo {volume} {24}} (\bibinfo {year}
  {2022})}\BibitemShut {NoStop}%
\bibitem [{\citenamefont {Wątorek}, \citenamefont {Kwapień},\ and\
  \citenamefont {Drożdż}(2023)}]{WatorekM-2023a}%
  \BibitemOpen
  \bibfield  {author} {\bibinfo {author} {\bibfnamefont {M.}~\bibnamefont
  {Wątorek}}, \bibinfo {author} {\bibfnamefont {J.}~\bibnamefont {Kwapień}},
  \ and\ \bibinfo {author} {\bibfnamefont {S.}~\bibnamefont {Drożdż}},\
  }\bibfield  {title} {\enquote {\bibinfo {title} {Cryptocurrencies are
  becoming part of the world global financial market},}\ }\href@noop {}
  {\bibfield  {journal} {\bibinfo  {journal} {Entropy}\ }\textbf {\bibinfo
  {volume} {25}},\ \bibinfo {pages} {377} (\bibinfo {year} {2023})}\BibitemShut
  {NoStop}%
\bibitem [{\citenamefont {Drożdż}, \citenamefont {Kwapień},\ and\
  \citenamefont {Wątorek}(2023)}]{Entropy2023}%
  \BibitemOpen
  \bibfield  {author} {\bibinfo {author} {\bibfnamefont {S.}~\bibnamefont
  {Drożdż}}, \bibinfo {author} {\bibfnamefont {J.}~\bibnamefont {Kwapień}},
  \ and\ \bibinfo {author} {\bibfnamefont {M.}~\bibnamefont {Wątorek}},\
  }\bibfield  {title} {\enquote {\bibinfo {title} {What is mature and what is
  still emerging in the cryptocurrency market?}}\ }\href@noop {} {\bibfield
  {journal} {\bibinfo  {journal} {Entropy}\ }\textbf {\bibinfo {volume} {25}}
  (\bibinfo {year} {2023})}\BibitemShut {NoStop}%
\bibitem [{\citenamefont {Makarov}\ and\ \citenamefont
  {Schoar}(2020)}]{MAKAROV2020}%
  \BibitemOpen
  \bibfield  {author} {\bibinfo {author} {\bibfnamefont {I.}~\bibnamefont
  {Makarov}}\ and\ \bibinfo {author} {\bibfnamefont {A.}~\bibnamefont
  {Schoar}},\ }\bibfield  {title} {\enquote {\bibinfo {title} {Trading and
  arbitrage in cryptocurrency markets},}\ }\href@noop {} {\bibfield  {journal}
  {\bibinfo  {journal} {Journal of Financial Economics}\ }\textbf {\bibinfo
  {volume} {135}},\ \bibinfo {pages} {293--319} (\bibinfo {year}
  {2020})}\BibitemShut {NoStop}%
\bibitem [{\citenamefont {Ante}(2023)}]{Ante2023}%
  \BibitemOpen
  \bibfield  {author} {\bibinfo {author} {\bibfnamefont {L.}~\bibnamefont
  {Ante}},\ }\bibfield  {title} {\enquote {\bibinfo {title} {How {Elon Musk's
  Twitter} activity moves cryptocurrency markets},}\ }\href@noop {} {\bibfield
  {journal} {\bibinfo  {journal} {Technological Forecasting and Social Change}\
  }\textbf {\bibinfo {volume} {186}},\ \bibinfo {pages} {122112} (\bibinfo
  {year} {2023})}\BibitemShut {NoStop}%
\bibitem [{\citenamefont {Li}, \citenamefont {Shin},\ and\ \citenamefont
  {Wang}(2021)}]{li2021cryptocurrency}%
  \BibitemOpen
  \bibfield  {author} {\bibinfo {author} {\bibfnamefont {T.}~\bibnamefont
  {Li}}, \bibinfo {author} {\bibfnamefont {D.}~\bibnamefont {Shin}}, \ and\
  \bibinfo {author} {\bibfnamefont {B.}~\bibnamefont {Wang}},\ }\href@noop {}
  {\enquote {\bibinfo {title} {Cryptocurrency pump-and-dump schemes},}\ }
  (\bibinfo {year} {2021}),\ \Eprint {http://arxiv.org/abs/3267041}
  {SSRN:3267041} \BibitemShut {NoStop}%
\bibitem [{\citenamefont {Dhawan}\ and\ \citenamefont
  {Putnins}(2022)}]{Dhawan2022}%
  \BibitemOpen
  \bibfield  {author} {\bibinfo {author} {\bibfnamefont {A.}~\bibnamefont
  {Dhawan}}\ and\ \bibinfo {author} {\bibfnamefont {T.~J.}\ \bibnamefont
  {Putnins}},\ }\bibfield  {title} {\enquote {\bibinfo {title} {A new wolf in
  town? pump-and-dump manipulation in cryptocurrency markets},}\ }\href@noop {}
  {\bibfield  {journal} {\bibinfo  {journal} {Review of Finance}\ }\textbf
  {\bibinfo {volume} {27}},\ \bibinfo {pages} {935--975} (\bibinfo {year}
  {2022})}\BibitemShut {NoStop}%
\bibitem [{\citenamefont {Cong}\ \emph {et~al.}(2022)\citenamefont {Cong},
  \citenamefont {Li}, \citenamefont {Tang},\ and\ \citenamefont
  {Yang}}]{cong2022crypto}%
  \BibitemOpen
  \bibfield  {author} {\bibinfo {author} {\bibfnamefont {L.~W.}\ \bibnamefont
  {Cong}}, \bibinfo {author} {\bibfnamefont {X.}~\bibnamefont {Li}}, \bibinfo
  {author} {\bibfnamefont {K.}~\bibnamefont {Tang}}, \ and\ \bibinfo {author}
  {\bibfnamefont {Y.}~\bibnamefont {Yang}},\ }\href@noop {} {\enquote {\bibinfo
  {title} {Crypto wash trading},}\ }\bibinfo {type} {Tech. Rep.}\ (\bibinfo
  {institution} {National Bureau of Economic Research},\ \bibinfo {year}
  {2022})\BibitemShut {NoStop}%
\bibitem [{\citenamefont {Baur}\ \emph {et~al.}(2019)\citenamefont {Baur},
  \citenamefont {Cahill}, \citenamefont {Godfrey},\ and\ \citenamefont
  {{(Frank) Liu}}}]{BAUR2019}%
  \BibitemOpen
  \bibfield  {author} {\bibinfo {author} {\bibfnamefont {D.~G.}\ \bibnamefont
  {Baur}}, \bibinfo {author} {\bibfnamefont {D.}~\bibnamefont {Cahill}},
  \bibinfo {author} {\bibfnamefont {K.}~\bibnamefont {Godfrey}}, \ and\
  \bibinfo {author} {\bibfnamefont {Z.}~\bibnamefont {{(Frank) Liu}}},\
  }\bibfield  {title} {\enquote {\bibinfo {title} {Bitcoin time-of-day,
  day-of-week and month-of-year effects in returns and trading volume},}\
  }\href@noop {} {\bibfield  {journal} {\bibinfo  {journal} {Finance Research
  Letters}\ }\textbf {\bibinfo {volume} {31}},\ \bibinfo {pages} {78--92}
  (\bibinfo {year} {2019})}\BibitemShut {NoStop}%
\bibitem [{\citenamefont {Catania}\ and\ \citenamefont
  {Sandholdt}(2019)}]{Catania2019}%
  \BibitemOpen
  \bibfield  {author} {\bibinfo {author} {\bibfnamefont {L.}~\bibnamefont
  {Catania}}\ and\ \bibinfo {author} {\bibfnamefont {M.}~\bibnamefont
  {Sandholdt}},\ }\bibfield  {title} {\enquote {\bibinfo {title} {Bitcoin at
  high frequency},}\ }\href@noop {} {\bibfield  {journal} {\bibinfo  {journal}
  {Journal of Risk and Financial Management}\ }\textbf {\bibinfo {volume} {12}}
  (\bibinfo {year} {2019})}\BibitemShut {NoStop}%
\bibitem [{\citenamefont {Dyhrberg}, \citenamefont {Foley},\ and\ \citenamefont
  {Svec}(2018)}]{Dyhrberg2018}%
  \BibitemOpen
  \bibfield  {author} {\bibinfo {author} {\bibfnamefont {A.~H.}\ \bibnamefont
  {Dyhrberg}}, \bibinfo {author} {\bibfnamefont {S.}~\bibnamefont {Foley}}, \
  and\ \bibinfo {author} {\bibfnamefont {J.}~\bibnamefont {Svec}},\ }\bibfield
  {title} {\enquote {\bibinfo {title} {How investible is {Bitcoin?} {Analyzing}
  the liquidity and transaction costs of {Bitcoin} markets},}\ }\href@noop {}
  {\bibfield  {journal} {\bibinfo  {journal} {Economics Letters}\ }\textbf
  {\bibinfo {volume} {171}},\ \bibinfo {pages} {140--143} (\bibinfo {year}
  {2018})}\BibitemShut {NoStop}%
\bibitem [{\citenamefont {Eross}\ \emph {et~al.}(2019)\citenamefont {Eross},
  \citenamefont {McGroarty}, \citenamefont {Urquhart},\ and\ \citenamefont
  {Wolfe}}]{Eross2019}%
  \BibitemOpen
  \bibfield  {author} {\bibinfo {author} {\bibfnamefont {A.}~\bibnamefont
  {Eross}}, \bibinfo {author} {\bibfnamefont {F.}~\bibnamefont {McGroarty}},
  \bibinfo {author} {\bibfnamefont {A.}~\bibnamefont {Urquhart}}, \ and\
  \bibinfo {author} {\bibfnamefont {S.}~\bibnamefont {Wolfe}},\ }\bibfield
  {title} {\enquote {\bibinfo {title} {The intraday dynamics of bitcoin},}\
  }\href@noop {} {\bibfield  {journal} {\bibinfo  {journal} {Research in
  International Business and Finance}\ }\textbf {\bibinfo {volume} {49}},\
  \bibinfo {pages} {71--81} (\bibinfo {year} {2019})}\BibitemShut {NoStop}%
\bibitem [{\citenamefont {Wang}, \citenamefont {Liu},\ and\ \citenamefont
  {Hsu}(2020)}]{Wang2020FRL}%
  \BibitemOpen
  \bibfield  {author} {\bibinfo {author} {\bibfnamefont {J.-N.}\ \bibnamefont
  {Wang}}, \bibinfo {author} {\bibfnamefont {H.-C.}\ \bibnamefont {Liu}}, \
  and\ \bibinfo {author} {\bibfnamefont {Y.-T.}\ \bibnamefont {Hsu}},\
  }\bibfield  {title} {\enquote {\bibinfo {title} {Time-of-day periodicities of
  trading volume and volatility in {Bitcoin} exchange: {Does} the stock market
  matter?}}\ }\href@noop {} {\bibfield  {journal} {\bibinfo  {journal} {Finance
  Research Letters}\ }\textbf {\bibinfo {volume} {34}},\ \bibinfo {pages}
  {101243} (\bibinfo {year} {2020})}\BibitemShut {NoStop}%
\bibitem [{\citenamefont {Omrane}, \citenamefont {Houidi},\ and\ \citenamefont
  {Savaser}(2023)}]{Omrane2023}%
  \BibitemOpen
  \bibfield  {author} {\bibinfo {author} {\bibfnamefont {W.~B.}\ \bibnamefont
  {Omrane}}, \bibinfo {author} {\bibfnamefont {F.}~\bibnamefont {Houidi}}, \
  and\ \bibinfo {author} {\bibfnamefont {T.}~\bibnamefont {Savaser}},\
  }\bibfield  {title} {\enquote {\bibinfo {title} {Macroeconomic news and
  intraday seasonal volatility in the cryptocurrency markets},}\ }\href@noop {}
  {\bibfield  {journal} {\bibinfo  {journal} {Applied Economics}\ }\textbf
  {\bibinfo {volume} {0}},\ \bibinfo {pages} {1--17} (\bibinfo {year}
  {2023})}\BibitemShut {NoStop}%
\bibitem [{\citenamefont {Hansen}, \citenamefont {Kim},\ and\ \citenamefont
  {Kimbrough}(2022)}]{Hansen2022}%
  \BibitemOpen
  \bibfield  {author} {\bibinfo {author} {\bibfnamefont {P.~R.}\ \bibnamefont
  {Hansen}}, \bibinfo {author} {\bibfnamefont {C.}~\bibnamefont {Kim}}, \ and\
  \bibinfo {author} {\bibfnamefont {W.}~\bibnamefont {Kimbrough}},\ }\bibfield
  {title} {\enquote {\bibinfo {title} {Periodicity in cryptocurrency volatility
  and liquidity},}\ }\href@noop {} {\bibfield  {journal} {\bibinfo  {journal}
  {Journal of Financial Econometrics}\ } (\bibinfo {year} {2022})}\BibitemShut
  {NoStop}%
\bibitem [{\citenamefont {Petukhina}, \citenamefont {Reule},\ and\
  \citenamefont {Härdle}(2021)}]{Petukhina2021}%
  \BibitemOpen
  \bibfield  {author} {\bibinfo {author} {\bibfnamefont {A.~A.}\ \bibnamefont
  {Petukhina}}, \bibinfo {author} {\bibfnamefont {R.~C.~G.}\ \bibnamefont
  {Reule}}, \ and\ \bibinfo {author} {\bibfnamefont {W.~K.}\ \bibnamefont
  {Härdle}},\ }\bibfield  {title} {\enquote {\bibinfo {title} {Rise of the
  machines? {I}ntraday high-frequency trading patterns of cryptocurrencies},}\
  }\href@noop {} {\bibfield  {journal} {\bibinfo  {journal} {The European
  Journal of Finance}\ }\textbf {\bibinfo {volume} {27}},\ \bibinfo {pages}
  {8--30} (\bibinfo {year} {2021})}\BibitemShut {NoStop}%
\bibitem [{\citenamefont {Aleti}\ and\ \citenamefont
  {Mizrach}(2021)}]{Aleti2021}%
  \BibitemOpen
  \bibfield  {author} {\bibinfo {author} {\bibfnamefont {S.}~\bibnamefont
  {Aleti}}\ and\ \bibinfo {author} {\bibfnamefont {B.}~\bibnamefont
  {Mizrach}},\ }\bibfield  {title} {\enquote {\bibinfo {title} {Bitcoin spot
  and futures market microstructure},}\ }\href@noop {} {\bibfield  {journal}
  {\bibinfo  {journal} {Journal of Futures Markets}\ }\textbf {\bibinfo
  {volume} {41}},\ \bibinfo {pages} {194--225} (\bibinfo {year}
  {2021})}\BibitemShut {NoStop}%
\bibitem [{\citenamefont {Dimpfl}\ and\ \citenamefont
  {Peter}(2021)}]{DIMPFL2021}%
  \BibitemOpen
  \bibfield  {author} {\bibinfo {author} {\bibfnamefont {T.}~\bibnamefont
  {Dimpfl}}\ and\ \bibinfo {author} {\bibfnamefont {F.~J.}\ \bibnamefont
  {Peter}},\ }\bibfield  {title} {\enquote {\bibinfo {title} {Nothing but
  noise? {Price} discovery across cryptocurrency exchanges},}\ }\href@noop {}
  {\bibfield  {journal} {\bibinfo  {journal} {Journal of Financial Markets}\
  }\textbf {\bibinfo {volume} {54}},\ \bibinfo {pages} {100584} (\bibinfo
  {year} {2021})}\BibitemShut {NoStop}%
\bibitem [{\citenamefont {Drożdż}\ \emph {et~al.}(2001)\citenamefont
  {Drożdż}, \citenamefont {Kwapień}, \citenamefont {Grümmer}, \citenamefont
  {Ruf},\ and\ \citenamefont {Speth}}]{Drozdz2001PhysA}%
  \BibitemOpen
  \bibfield  {author} {\bibinfo {author} {\bibfnamefont {S.}~\bibnamefont
  {Drożdż}}, \bibinfo {author} {\bibfnamefont {J.}~\bibnamefont {Kwapień}},
  \bibinfo {author} {\bibfnamefont {F.}~\bibnamefont {Grümmer}}, \bibinfo
  {author} {\bibfnamefont {F.}~\bibnamefont {Ruf}}, \ and\ \bibinfo {author}
  {\bibfnamefont {J.}~\bibnamefont {Speth}},\ }\bibfield  {title} {\enquote
  {\bibinfo {title} {Quantifying the dynamics of financial correlations},}\
  }\href@noop {} {\bibfield  {journal} {\bibinfo  {journal} {Physica A}\
  }\textbf {\bibinfo {volume} {299}},\ \bibinfo {pages} {144--153} (\bibinfo
  {year} {2001})}\BibitemShut {NoStop}%
\bibitem [{\citenamefont {Kwapie\ifmmode~\acute{n}\else \'{n}\fi{}},
  \citenamefont {Dro\ifmmode \dot{z}\else \.{z}\fi{}d\ifmmode~\dot{z}\else
  \.{z}\fi{}},\ and\ \citenamefont {Ioannides}(2000)}]{Kwapien2000}%
  \BibitemOpen
  \bibfield  {author} {\bibinfo {author} {\bibfnamefont {J.}~\bibnamefont
  {Kwapie\ifmmode~\acute{n}\else \'{n}\fi{}}}, \bibinfo {author} {\bibfnamefont
  {S.}~\bibnamefont {Dro\ifmmode \dot{z}\else \.{z}\fi{}d\ifmmode~\dot{z}\else
  \.{z}\fi{}}}, \ and\ \bibinfo {author} {\bibfnamefont {A.~A.}\ \bibnamefont
  {Ioannides}},\ }\bibfield  {title} {\enquote {\bibinfo {title} {Temporal
  correlations versus noise in the correlation matrix formalism: An example of
  the brain auditory response},}\ }\href@noop {} {\bibfield  {journal}
  {\bibinfo  {journal} {Physical Review E}\ }\textbf {\bibinfo {volume} {62}},\
  \bibinfo {pages} {5557--5564} (\bibinfo {year} {2000})}\BibitemShut {NoStop}%
\bibitem [{\citenamefont {Dro\.zd\.z}\ \emph {et~al.}(2000)\citenamefont
  {Dro\.zd\.z}, \citenamefont {G\"ummer}, \citenamefont {G\'orski},
  \citenamefont {Ruf},\ and\ \citenamefont {Speth}}]{Drozdz2000}%
  \BibitemOpen
  \bibfield  {author} {\bibinfo {author} {\bibfnamefont {S.}~\bibnamefont
  {Dro\.zd\.z}}, \bibinfo {author} {\bibfnamefont {F.}~\bibnamefont
  {G\"ummer}}, \bibinfo {author} {\bibfnamefont {A.~Z.}\ \bibnamefont
  {G\'orski}}, \bibinfo {author} {\bibfnamefont {F.}~\bibnamefont {Ruf}}, \
  and\ \bibinfo {author} {\bibfnamefont {J.}~\bibnamefont {Speth}},\ }\bibfield
   {title} {\enquote {\bibinfo {title} {Dynamics of competition between
  collectivity and noise in the stock market},}\ }\href {\doibase
  10.1016/S0378-4371(00)00383-6} {\bibfield  {journal} {\bibinfo  {journal}
  {Physica A}\ }\textbf {\bibinfo {volume} {287}},\ \bibinfo {pages} {440--449}
  (\bibinfo {year} {2000})}\BibitemShut {NoStop}%
\bibitem [{\citenamefont {James}\ and\ \citenamefont
  {Menzies}(2021)}]{James2021chaos}%
  \BibitemOpen
  \bibfield  {author} {\bibinfo {author} {\bibfnamefont {N.}~\bibnamefont
  {James}}\ and\ \bibinfo {author} {\bibfnamefont {M.}~\bibnamefont
  {Menzies}},\ }\bibfield  {title} {\enquote {\bibinfo {title} {{Efficiency of
  communities and financial markets during the 2020 pandemic}},}\ }\href@noop
  {} {\bibfield  {journal} {\bibinfo  {journal} {Chaos}\ }\textbf {\bibinfo
  {volume} {31}},\ \bibinfo {pages} {083116} (\bibinfo {year}
  {2021})}\BibitemShut {NoStop}%
\bibitem [{\citenamefont {James}, \citenamefont {Menzies},\ and\ \citenamefont
  {Gottwald}(2022)}]{James2022PhysAstock}%
  \BibitemOpen
  \bibfield  {author} {\bibinfo {author} {\bibfnamefont {N.}~\bibnamefont
  {James}}, \bibinfo {author} {\bibfnamefont {M.}~\bibnamefont {Menzies}}, \
  and\ \bibinfo {author} {\bibfnamefont {G.~A.}\ \bibnamefont {Gottwald}},\
  }\bibfield  {title} {\enquote {\bibinfo {title} {On financial market
  correlation structures and diversification benefits across and within equity
  sectors},}\ }\href@noop {} {\bibfield  {journal} {\bibinfo  {journal}
  {Physica A}\ }\textbf {\bibinfo {volume} {604}},\ \bibinfo {pages} {127682}
  (\bibinfo {year} {2022})}\BibitemShut {NoStop}%
\bibitem [{\citenamefont {Chaudhari}\ and\ \citenamefont
  {Crane}(2020)}]{Chaudhari2020}%
  \BibitemOpen
  \bibfield  {author} {\bibinfo {author} {\bibfnamefont {H.}~\bibnamefont
  {Chaudhari}}\ and\ \bibinfo {author} {\bibfnamefont {M.}~\bibnamefont
  {Crane}},\ }\bibfield  {title} {\enquote {\bibinfo {title} {Cross-correlation
  dynamics and community structures of cryptocurrencies},}\ }\href@noop {}
  {\bibfield  {journal} {\bibinfo  {journal} {Journal of Computational
  Science}\ }\textbf {\bibinfo {volume} {44}},\ \bibinfo {pages} {101130}
  (\bibinfo {year} {2020})}\BibitemShut {NoStop}%
\bibitem [{\citenamefont {Dro\.{z}d\.{z}}\ \emph {et~al.}(2020)\citenamefont
  {Dro\.{z}d\.{z}}, \citenamefont {Minati}, \citenamefont {O\'swi\k{e}cimka},
  \citenamefont {Stanuszek},\ and\ \citenamefont {W\k{a}torek}}]{Chaos2020}%
  \BibitemOpen
  \bibfield  {author} {\bibinfo {author} {\bibfnamefont {S.}~\bibnamefont
  {Dro\.{z}d\.{z}}}, \bibinfo {author} {\bibfnamefont {L.}~\bibnamefont
  {Minati}}, \bibinfo {author} {\bibfnamefont {P.}~\bibnamefont
  {O\'swi\k{e}cimka}}, \bibinfo {author} {\bibfnamefont {M.}~\bibnamefont
  {Stanuszek}}, \ and\ \bibinfo {author} {\bibfnamefont {M.}~\bibnamefont
  {W\k{a}torek}},\ }\bibfield  {title} {\enquote {\bibinfo {title} {Competition
  of noise and collectivity in global cryptocurrency trading: {R}oute to a
  self-contained market},}\ }\href@noop {} {\bibfield  {journal} {\bibinfo
  {journal} {Chaos}\ }\textbf {\bibinfo {volume} {30}},\ \bibinfo {pages}
  {023122} (\bibinfo {year} {2020})}\BibitemShut {NoStop}%
\bibitem [{\citenamefont {James}\ and\ \citenamefont
  {Menzies}(2022)}]{James2022}%
  \BibitemOpen
  \bibfield  {author} {\bibinfo {author} {\bibfnamefont {N.}~\bibnamefont
  {James}}\ and\ \bibinfo {author} {\bibfnamefont {M.}~\bibnamefont
  {Menzies}},\ }\bibfield  {title} {\enquote {\bibinfo {title} {Collective
  correlations, dynamics, and behavioural inconsistencies of the cryptocurrency
  market over time},}\ }\href@noop {} {\bibfield  {journal} {\bibinfo
  {journal} {Nonlinear Dynamics}\ }\textbf {\bibinfo {volume} {107}},\ \bibinfo
  {pages} {4001--4017} (\bibinfo {year} {2022})}\BibitemShut {NoStop}%
\bibitem [{\citenamefont {Nguyen}\ \emph {et~al.}(2022)\citenamefont {Nguyen},
  \citenamefont {Mai}, \citenamefont {Bezbradica},\ and\ \citenamefont
  {Crane}}]{Nguyen2022}%
  \BibitemOpen
  \bibfield  {author} {\bibinfo {author} {\bibfnamefont {A.~P.~N.}\
  \bibnamefont {Nguyen}}, \bibinfo {author} {\bibfnamefont {T.~T.}\
  \bibnamefont {Mai}}, \bibinfo {author} {\bibfnamefont {M.}~\bibnamefont
  {Bezbradica}}, \ and\ \bibinfo {author} {\bibfnamefont {M.}~\bibnamefont
  {Crane}},\ }\bibfield  {title} {\enquote {\bibinfo {title} {The
  cryptocurrency market in transition before and after {COVID-19}: {An}
  opportunity for investors?}}\ }\href@noop {} {\bibfield  {journal} {\bibinfo
  {journal} {Entropy}\ }\textbf {\bibinfo {volume} {24}} (\bibinfo {year}
  {2022})}\BibitemShut {NoStop}%
\bibitem [{\citenamefont {Gavin}\ and\ \citenamefont
  {Crane}(2023)}]{gavin2023community}%
  \BibitemOpen
  \bibfield  {author} {\bibinfo {author} {\bibfnamefont {J.}~\bibnamefont
  {Gavin}}\ and\ \bibinfo {author} {\bibfnamefont {M.}~\bibnamefont {Crane}},\
  }\bibfield  {title} {\enquote {\bibinfo {title} {Community detection in
  cryptocurrencies with potential applications to portfolio diversification},}\
  }in\ \href@noop {} {\emph {\bibinfo {booktitle} {FinTech Research and
  Applications: Challenges and Opportunities}}}\ (\bibinfo  {publisher} {World
  Scientific},\ \bibinfo {year} {2023})\ pp.\ \bibinfo {pages}
  {177--202}\BibitemShut {NoStop}%
\bibitem [{\citenamefont {James}\ and\ \citenamefont
  {Menzies}(2023)}]{James2023}%
  \BibitemOpen
  \bibfield  {author} {\bibinfo {author} {\bibfnamefont {N.}~\bibnamefont
  {James}}\ and\ \bibinfo {author} {\bibfnamefont {M.}~\bibnamefont
  {Menzies}},\ }\bibfield  {title} {\enquote {\bibinfo {title} {Collective
  dynamics, diversification and optimal portfolio construction for
  cryptocurrencies},}\ }\href@noop {} {\bibfield  {journal} {\bibinfo
  {journal} {Entropy}\ }\textbf {\bibinfo {volume} {25}} (\bibinfo {year}
  {2023})}\BibitemShut {NoStop}%
\bibitem [{\citenamefont {Kwapie\'n}, \citenamefont {W\k{a}torek},\ and\
  \citenamefont {Dro\.{z}d\.{z}}(2021)}]{entropy2021b}%
  \BibitemOpen
  \bibfield  {author} {\bibinfo {author} {\bibfnamefont {J.}~\bibnamefont
  {Kwapie\'n}}, \bibinfo {author} {\bibfnamefont {M.}~\bibnamefont
  {W\k{a}torek}}, \ and\ \bibinfo {author} {\bibfnamefont {S.}~\bibnamefont
  {Dro\.{z}d\.{z}}},\ }\bibfield  {title} {\enquote {\bibinfo {title}
  {Cryptocurrency market consolidation in 2020-2021},}\ }\href@noop {}
  {\bibfield  {journal} {\bibinfo  {journal} {Entropy}\ }\textbf {\bibinfo
  {volume} {23}} (\bibinfo {year} {2021})}\BibitemShut {NoStop}%
\bibitem [{\citenamefont {James}(2022)}]{JAMES2022PhysD}%
  \BibitemOpen
  \bibfield  {author} {\bibinfo {author} {\bibfnamefont {N.}~\bibnamefont
  {James}},\ }\bibfield  {title} {\enquote {\bibinfo {title} {Evolutionary
  correlation, regime switching, spectral dynamics and optimal trading
  strategies for cryptocurrencies and equities},}\ }\href@noop {} {\bibfield
  {journal} {\bibinfo  {journal} {Physica D}\ }\textbf {\bibinfo {volume}
  {434}},\ \bibinfo {pages} {133262} (\bibinfo {year} {2022})}\BibitemShut
  {NoStop}%
\bibitem [{Binance()}]{Binance}%
  \BibitemOpen
  Binance,\ \href@noop {} {\enquote {\bibinfo {title} {Binance},}\ }\bibinfo
  {howpublished} {\url{https://www.binance.com/}}\BibitemShut {NoStop}%
\bibitem [{\citenamefont {W\k{a}torek}, \citenamefont {Kwapie\'n},\ and\
  \citenamefont {Dro\.{z}d\.{z}}(2022)}]{watorekfutnet2022}%
  \BibitemOpen
  \bibfield  {author} {\bibinfo {author} {\bibfnamefont {M.}~\bibnamefont
  {W\k{a}torek}}, \bibinfo {author} {\bibfnamefont {J.}~\bibnamefont
  {Kwapie\'n}}, \ and\ \bibinfo {author} {\bibfnamefont {S.}~\bibnamefont
  {Dro\.{z}d\.{z}}},\ }\bibfield  {title} {\enquote {\bibinfo {title}
  {Multifractal cross-correlations of bitcoin and ether trading characteristics
  in the post-{COVID-19} time},}\ }\href@noop {} {\bibfield  {journal}
  {\bibinfo  {journal} {Future Internet}\ }\textbf {\bibinfo {volume} {14}}
  (\bibinfo {year} {2022})}\BibitemShut {NoStop}%
\bibitem [{\citenamefont {Nani}(2022)}]{Nani2022}%
  \BibitemOpen
  \bibfield  {author} {\bibinfo {author} {\bibfnamefont {A.}~\bibnamefont
  {Nani}},\ }\bibfield  {title} {\enquote {\bibinfo {title} {The doge worth 88
  billion dollars: A case study of {Dogecoin}},}\ }\href@noop {} {\bibfield
  {journal} {\bibinfo  {journal} {Convergence}\ }\textbf {\bibinfo {volume}
  {28}},\ \bibinfo {pages} {1719--1736} (\bibinfo {year} {2022})}\BibitemShut
  {NoStop}%
\bibitem [{\citenamefont {Muravyev}\ and\ \citenamefont
  {Picard}(2022)}]{Muravyev2022}%
  \BibitemOpen
  \bibfield  {author} {\bibinfo {author} {\bibfnamefont {D.}~\bibnamefont
  {Muravyev}}\ and\ \bibinfo {author} {\bibfnamefont {J.}~\bibnamefont
  {Picard}},\ }\bibfield  {title} {\enquote {\bibinfo {title} {Does trade
  clustering reduce trading costs? evidence from periodicity in algorithmic
  trading},}\ }\href@noop {} {\bibfield  {journal} {\bibinfo  {journal}
  {Financial Management}\ }\textbf {\bibinfo {volume} {51}},\ \bibinfo {pages}
  {1201--1229} (\bibinfo {year} {2022})}\BibitemShut {NoStop}%
\bibitem [{\citenamefont {Broussard}\ and\ \citenamefont
  {Nikiforov}(2014)}]{Broussard2014}%
  \BibitemOpen
  \bibfield  {author} {\bibinfo {author} {\bibfnamefont {J.~P.}\ \bibnamefont
  {Broussard}}\ and\ \bibinfo {author} {\bibfnamefont {A.}~\bibnamefont
  {Nikiforov}},\ }\bibfield  {title} {\enquote {\bibinfo {title} {Intraday
  periodicity in algorithmic trading},}\ }\href@noop {} {\bibfield  {journal}
  {\bibinfo  {journal} {Journal of International Financial Markets,
  Institutions and Money}\ }\textbf {\bibinfo {volume} {30}},\ \bibinfo {pages}
  {196--204} (\bibinfo {year} {2014})}\BibitemShut {NoStop}%
\bibitem [{Similarweb()}]{similarweb}%
  \BibitemOpen
  Similarweb,\ \href@noop {} {\enquote {\bibinfo {title} {{Similarweb}},}\
  }\bibinfo {howpublished}
  {\url{https://www.similarweb.com/website/binance.com/}}\BibitemShut {NoStop}%
\bibitem [{\citenamefont {Mehta}(2004)}]{randommatrix}%
  \BibitemOpen
  \bibfield  {author} {\bibinfo {author} {\bibfnamefont {M.~L.}\ \bibnamefont
  {Mehta}},\ }\href@noop {} {\emph {\bibinfo {title} {Random Matrices}}}\
  (\bibinfo  {publisher} {Elsevier},\ \bibinfo {year} {2004})\BibitemShut
  {NoStop}%
\bibitem [{\citenamefont {Wishart}(1928)}]{wishart1928}%
  \BibitemOpen
  \bibfield  {author} {\bibinfo {author} {\bibfnamefont {J.}~\bibnamefont
  {Wishart}},\ }\bibfield  {title} {\enquote {\bibinfo {title} {The generalised
  product moment distribution in samples from a normal multivariate
  population},}\ }\href@noop {} {\bibfield  {journal} {\bibinfo  {journal}
  {Biometrika}\ }\textbf {\bibinfo {volume} {20A}},\ \bibinfo {pages} {32--52}
  (\bibinfo {year} {1928})}\BibitemShut {NoStop}%
\bibitem [{\citenamefont {Mar{\v{c}}enko}\ and\ \citenamefont
  {Pastur}(1967)}]{Marchenko1967}%
  \BibitemOpen
  \bibfield  {author} {\bibinfo {author} {\bibfnamefont {V.~A.}\ \bibnamefont
  {Mar{\v{c}}enko}}\ and\ \bibinfo {author} {\bibfnamefont {L.~A.}\
  \bibnamefont {Pastur}},\ }\bibfield  {title} {\enquote {\bibinfo {title}
  {Distribution of eigenvalues for some sets of random matrices},}\ }\href
  {\doibase 10.1070/sm1967v001n04abeh001994} {\bibfield  {journal} {\bibinfo
  {journal} {Mathematics of the {USSR-Sbornik}}\ }\textbf {\bibinfo {volume}
  {1}},\ \bibinfo {pages} {457--483} (\bibinfo {year} {1967})}\BibitemShut
  {NoStop}%
\bibitem [{\citenamefont {Gopikrishnan}\ \emph {et~al.}(1999)\citenamefont
  {Gopikrishnan}, \citenamefont {Plerou}, \citenamefont {Nunes~Amaral},
  \citenamefont {Meyer},\ and\ \citenamefont {Stanley}}]{Gopikrishnan1999}%
  \BibitemOpen
  \bibfield  {author} {\bibinfo {author} {\bibfnamefont {P.}~\bibnamefont
  {Gopikrishnan}}, \bibinfo {author} {\bibfnamefont {V.}~\bibnamefont
  {Plerou}}, \bibinfo {author} {\bibfnamefont {L.~A.}\ \bibnamefont
  {Nunes~Amaral}}, \bibinfo {author} {\bibfnamefont {M.}~\bibnamefont {Meyer}},
  \ and\ \bibinfo {author} {\bibfnamefont {H.~E.}\ \bibnamefont {Stanley}},\
  }\bibfield  {title} {\enquote {\bibinfo {title} {Scaling of the distribution
  of fluctuations of financial market indices},}\ }\href@noop {} {\bibfield
  {journal} {\bibinfo  {journal} {Physical Review E}\ }\textbf {\bibinfo
  {volume} {60}},\ \bibinfo {pages} {5305--5316} (\bibinfo {year}
  {1999})}\BibitemShut {NoStop}%
\bibitem [{\citenamefont {Ausloos}(2000)}]{Ausloos2000}%
  \BibitemOpen
  \bibfield  {author} {\bibinfo {author} {\bibfnamefont {M.}~\bibnamefont
  {Ausloos}},\ }\bibfield  {title} {\enquote {\bibinfo {title} {Statistical
  physics in foreign exchange currency and stock markets},}\ }\href@noop {}
  {\bibfield  {journal} {\bibinfo  {journal} {Physica A}\ }\textbf {\bibinfo
  {volume} {285}},\ \bibinfo {pages} {48--65} (\bibinfo {year}
  {2000})}\BibitemShut {NoStop}%
\bibitem [{\citenamefont {Cont}(2001)}]{ContR-2001a}%
  \BibitemOpen
  \bibfield  {author} {\bibinfo {author} {\bibfnamefont {R.}~\bibnamefont
  {Cont}},\ }\bibfield  {title} {\enquote {\bibinfo {title} {Empirical
  properties of asset returns: stylized facts and statistical issues},}\ }\href
  {http://www.cmap.polytechnique.fr/} {\bibfield  {journal} {\bibinfo
  {journal} {Quantitative Finance}\ }\textbf {\bibinfo {volume} {1}},\ \bibinfo
  {pages} {223--236} (\bibinfo {year} {2001})}\BibitemShut {NoStop}%
\bibitem [{\citenamefont {Jiang}\ \emph {et~al.}(2019)\citenamefont {Jiang},
  \citenamefont {Xie}, \citenamefont {Zhou},\ and\ \citenamefont
  {Sornette}}]{jiang2019multifractal}%
  \BibitemOpen
  \bibfield  {author} {\bibinfo {author} {\bibfnamefont {Z.-Q.}\ \bibnamefont
  {Jiang}}, \bibinfo {author} {\bibfnamefont {W.-J.}\ \bibnamefont {Xie}},
  \bibinfo {author} {\bibfnamefont {W.-X.}\ \bibnamefont {Zhou}}, \ and\
  \bibinfo {author} {\bibfnamefont {D.}~\bibnamefont {Sornette}},\ }\bibfield
  {title} {\enquote {\bibinfo {title} {Multifractal analysis of financial
  markets: a review},}\ }\href@noop {} {\bibfield  {journal} {\bibinfo
  {journal} {Reports on Progress in Physics}\ }\textbf {\bibinfo {volume}
  {82}},\ \bibinfo {pages} {125901} (\bibinfo {year} {2019})}\BibitemShut
  {NoStop}%
\bibitem [{\citenamefont {Klamut}\ \emph {et~al.}(2020)\citenamefont {Klamut},
  \citenamefont {Kutner}, \citenamefont {Gubiec},\ and\ \citenamefont
  {Struzik}}]{Klamut2020}%
  \BibitemOpen
  \bibfield  {author} {\bibinfo {author} {\bibfnamefont {J.}~\bibnamefont
  {Klamut}}, \bibinfo {author} {\bibfnamefont {R.}~\bibnamefont {Kutner}},
  \bibinfo {author} {\bibfnamefont {T.}~\bibnamefont {Gubiec}}, \ and\ \bibinfo
  {author} {\bibfnamefont {Z.~R.}\ \bibnamefont {Struzik}},\ }\bibfield
  {title} {\enquote {\bibinfo {title} {Multibranch multifractality and the
  phase transitions in time series of mean interevent times},}\ }\href@noop {}
  {\bibfield  {journal} {\bibinfo  {journal} {Physical Review E}\ }\textbf
  {\bibinfo {volume} {101}},\ \bibinfo {pages} {063303} (\bibinfo {year}
  {2020})}\BibitemShut {NoStop}%
\bibitem [{\citenamefont {Klamut}\ and\ \citenamefont
  {Gubiec}(2021)}]{Klamut2021}%
  \BibitemOpen
  \bibfield  {author} {\bibinfo {author} {\bibfnamefont {J.}~\bibnamefont
  {Klamut}}\ and\ \bibinfo {author} {\bibfnamefont {T.}~\bibnamefont
  {Gubiec}},\ }\bibfield  {title} {\enquote {\bibinfo {title} {Continuous time
  random walk with correlated waiting times. the crucial role of inter-trade
  times in volatility clustering},}\ }\href@noop {} {\bibfield  {journal}
  {\bibinfo  {journal} {Entropy}\ }\textbf {\bibinfo {volume} {23}} (\bibinfo
  {year} {2021})}\BibitemShut {NoStop}%
\bibitem [{\citenamefont {Drożdż}\ \emph {et~al.}(2002)\citenamefont
  {Drożdż}, \citenamefont {Kwapień}, \citenamefont {Speth},\ and\
  \citenamefont {Wójcik}}]{Drozdz2002RMT}%
  \BibitemOpen
  \bibfield  {author} {\bibinfo {author} {\bibfnamefont {S.}~\bibnamefont
  {Drożdż}}, \bibinfo {author} {\bibfnamefont {J.}~\bibnamefont {Kwapień}},
  \bibinfo {author} {\bibfnamefont {J.}~\bibnamefont {Speth}}, \ and\ \bibinfo
  {author} {\bibfnamefont {M.}~\bibnamefont {Wójcik}},\ }\bibfield  {title}
  {\enquote {\bibinfo {title} {Identifying complexity by means of matrices},}\
  }\href@noop {} {\bibfield  {journal} {\bibinfo  {journal} {Physica A}\
  }\textbf {\bibinfo {volume} {314}},\ \bibinfo {pages} {355--361} (\bibinfo
  {year} {2002})}\BibitemShut {NoStop}%
\bibitem [{\citenamefont {Bouchaud}(2010)}]{BouchaudJP-2010a}%
  \BibitemOpen
  \bibfield  {author} {\bibinfo {author} {\bibfnamefont {J.-P.}\ \bibnamefont
  {Bouchaud}},\ }\bibfield  {title} {\enquote {\bibinfo {title} {Price
  impact},}\ }in\ \href {http://arxiv.org/abs/0903.2428} {\emph {\bibinfo
  {booktitle} {Encyclopedia of Quantitative Finance}}}\ (\bibinfo  {publisher}
  {Cambridge University Press},\ \bibinfo {year} {2010})\ pp.\ \bibinfo {pages}
  {1--6}\BibitemShut {NoStop}%
\bibitem [{\citenamefont {Shahzad}, \citenamefont {Anas},\ and\ \citenamefont
  {Bouri}(2022)}]{SHAHZAD2022Doge}%
  \BibitemOpen
  \bibfield  {author} {\bibinfo {author} {\bibfnamefont {S.~J.~H.}\
  \bibnamefont {Shahzad}}, \bibinfo {author} {\bibfnamefont {M.}~\bibnamefont
  {Anas}}, \ and\ \bibinfo {author} {\bibfnamefont {E.}~\bibnamefont {Bouri}},\
  }\bibfield  {title} {\enquote {\bibinfo {title} {Price explosiveness in
  cryptocurrencies and {Elon Musk's} tweets},}\ }\href@noop {} {\bibfield
  {journal} {\bibinfo  {journal} {Finance Research Letters}\ }\textbf {\bibinfo
  {volume} {47}},\ \bibinfo {pages} {102695} (\bibinfo {year}
  {2022})}\BibitemShut {NoStop}%
\bibitem [{\citenamefont {James}, \citenamefont {Menzies},\ and\ \citenamefont
  {Chin}(2022)}]{James2022inf}%
  \BibitemOpen
  \bibfield  {author} {\bibinfo {author} {\bibfnamefont {N.}~\bibnamefont
  {James}}, \bibinfo {author} {\bibfnamefont {M.}~\bibnamefont {Menzies}}, \
  and\ \bibinfo {author} {\bibfnamefont {K.}~\bibnamefont {Chin}},\ }\bibfield
  {title} {\enquote {\bibinfo {title} {Economic state classification and
  portfolio optimisation with application to stagflationary environments},}\
  }\href@noop {} {\bibfield  {journal} {\bibinfo  {journal} {Chaos, Solitons \&
  Fractals}\ }\textbf {\bibinfo {volume} {164}},\ \bibinfo {pages} {112664}
  (\bibinfo {year} {2022})}\BibitemShut {NoStop}%
\bibitem [{\citenamefont {Choi}\ and\ \citenamefont {Shin}(2022)}]{Choi2022}%
  \BibitemOpen
  \bibfield  {author} {\bibinfo {author} {\bibfnamefont {S.}~\bibnamefont
  {Choi}}\ and\ \bibinfo {author} {\bibfnamefont {J.}~\bibnamefont {Shin}},\
  }\bibfield  {title} {\enquote {\bibinfo {title} {Bitcoin: {An} inflation
  hedge but not a safe haven},}\ }\href@noop {} {\bibfield  {journal} {\bibinfo
   {journal} {Finance Research Letters}\ }\textbf {\bibinfo {volume} {46}},\
  \bibinfo {pages} {102379} (\bibinfo {year} {2022})}\BibitemShut {NoStop}%
\end{thebibliography}%

\end{document}